\documentclass[11pt]{article}
\usepackage[totalwidth=460truept,totalheight=600truept]{geometry}
\usepackage{epsfig,latexsym,amssymb}

\def\theequation{\arabic{section}.\arabic{equation}}

\renewcommand{\theequation}{\thesection.\arabic{equation}}
\global\arraycolsep=1truept
\input epsf
\input{tcilatex}

\begin{document}

\hfill \hfill IFUP-TH 2006/24

\vskip 1.4truecm

\begin{center}
{\huge \textbf{Renormalizable Acausal Theories Of\ }}\vskip .5truecm{\huge 
\textbf{\ Classical Gravity Coupled With }}\vskip .5truecm{\huge \textbf{\
Interacting Quantum Fields}} \vskip 1.5truecm

\textsl{Damiano Anselmi and Milenko Halat}

\textit{Dipartimento di Fisica ``Enrico Fermi'', Universit\`{a} di Pisa, }

\textit{Largo Pontecorvo 3, I-56127 Pisa, Italy, }

\textit{and INFN, Sezione di Pisa, Pisa, Italy}

damiano.anselmi@df.unipi.it, milenko.halat@df.unipi.it
\end{center}

\vskip 2truecm

\begin{center}
\textbf{Abstract}
\end{center}

\bigskip

{\small We prove the renormalizability of various theories of classical
gravity coupled with interacting quantum fields. The models contain vertices
with dimensionalities greater than four, a finite number of matter operators
and a finite or reduced number of independent couplings.\ An interesting
class of models is obtained from ordinary power-counting renormalizable
theories, letting the couplings depend on the scalar curvature }$R${\small \
of spacetime. The divergences are removed without introducing
higher-derivative kinetic terms in the gravitational sector. The metric
tensor has a non-trivial running, even if it is not quantized. The results
are proved applying a certain map that converts classical instabilities, due
to higher derivatives, into classical violations of causality, whose effects
become observable at sufficiently high energies. We study acausal
Einstein-Yang-Mills theory with an }$R${\small -dependent gauge coupling in
detail. We derive all-order formulas for the beta functions of the
dimensionality-six gravitational vertices induced by renormalization, such
as }$RR_{\mu \nu \rho \sigma }R^{\mu \nu \rho \sigma }${\small . Such beta
functions are related to the trace-anomaly coefficients of the matter
subsector.}

\vskip 1truecm

\vfill\eject

\section{Introduction}

\setcounter{equation}{0}

The renormalization of quantum field theory in curved space has been widely
studied \cite{birreldavis}. Treating the metric tensor as a $c$-number and
neglecting its quantum fluctuations, classical gravity coupled with
quantized fields can be used as a low-energy effective field theory, to
include the radiative corrections to the Einstein field equations generated
by the matter fields circulating in the loops. Moveover, it provides an
interesting arena and a laboratory to test ideas about renormalizability
beyond power counting. Although there exist persuasive reasons to believe
that gravity must be quantized, definitive theoretical arguments and
experimental proofs are still missing \cite
{bohr,moller,feynman,eppley,mattingly}. Thus it is meaningful to study the
physical consequences of the assumption that classical gravity coupled with
quantized fields is a fundamental theory, valid at arbitrarily high
energies, instead of just an effective one.

When matter is embedded in a curved background, renormalization generates
the gravitational counterterms $R_{\mu \nu }R^{\mu \nu }$ and $R^{2}$ \cite
{thooftveltman}. Such counterterms can be renormalized in two ways. One
possibility is to add the same terms to the lagrangian, if they are not
already present, multiplied by independent couplings, and reabsorb the
divergences into those couplings. The resulting theory is higher-derivative.
Expanding the metric tensor around flat space, the lagrangian contains
higher-derivative kinetic terms, which are responsible for instabilities at
the classical level and violations of unitarity at the quantum level. In
particular, higher-derivative quantum gravity is renormalizable, but not
unitary \cite{fradkin}: the propagator falls off sufficiently rapidly at
high energies to ensure power-counting renormalizability, but propagates
ghosts. An alternative way to remove counterterms, that applies only when
they have an appropriate form, is to use field redefinitions. Applied to $%
R_{\mu \nu }R^{\mu \nu }$ and $R^{2}$, field redefinitions can convert the
undesirable higher-derivative kinetic terms into new types of vertices that
couple gravity to matter.

In a recent paper \cite{clgr} it was shown that in classical gravity coupled
with quantum matter the second method of subtraction can be consistently
implemented to all orders in the perturbative expansion, redefining the
metric tensor by means of a certain map $\mathcal{M}$. Since the terms $%
R_{\mu \nu }R^{\mu \nu }$ and $R^{2}$ are proportional to the vacuum field
equations, a metric redefinition $g_{\mu \nu }^{\prime }=g_{\mu \nu }+\delta
g_{\mu \nu }$ can obviously reabsorb them into the Einstein term to the
first order in $\delta g_{\mu \nu }$. In the presence of matter the
redefinition generates vertices that couple the matter stress-tensor to the
Ricci tensor. The map $\mathcal{M}$ promotes such a field redefinition to
all orders. This is possible because $R_{\mu \nu }R^{\mu \nu }$ and $R^{2}$
are not only proportional, but also \textit{quadratically} proportional to
the Einstein vacuum field equations. If gravity is classical, the map $%
\mathcal{M}$ preserves the renormalizability of the theory. However, while
the map $\mathcal{M}$ eliminates the higher-derivative kinetic terms, and
therefore the instabilities, in general it produces classical violations of
causality, detectable in principle at high energies.

In this paper we prove the renormalizability of new types of acausal
theories of classical gravity coupled with quantized fields. In a class of
theories the matter sector contains all composite operators that have
dimensionalities smaller than or equal to four and the gravitational sector
contains arbitrary functions of the metric, therefore an infinite number of
independent couplings. The arbitrariness of such theories can be reduced,
preserving the renormalizability, by appropriate \textit{reductions of
couplings}. Specifically, we prove the renormalizability of the models that
are obtained from ordinary power-counting renormalizable theories embedded
in curved space, when the couplings are allowed to depend on the scalar
curvature $R$ of spacetime. In these models the arbitrariness is reduced to
a few functions of $R$. Every models has a higher-derivative version and an
acausal version, and we can switch between the two using the map $\mathcal{M}
$.

In particular, we study acausal Einstein-Yang-Mills theory with an $R$%
-dependent gauge coupling. Renormalization induces new pure gravitational
vertices, such as $RR_{\mu \nu \rho \sigma }R^{\mu \nu \rho \sigma }$. We
work out exact formulas for the beta functions of the dimensionality-six
gravitational vertices, which are related to the trace-anomaly coefficients
of the matter subsector.

Our arguments, in particular the map $\mathcal{M}$, do not extend to quantum
gravity. Nevertheless, our techniques and results might have some impact
also on the task of quantizing gravity.

\bigskip

The paper is organized as follows. In section 2 we recall the results of 
\cite{clgr} that are used here. In section 3 we apply them to describe
acausal Einstein-Yang-Mills theory explicitly. In section 4 we prove a first
generalization of the results of \cite{clgr}, proving the renormalizability
of a more general class of theories, but with essentially the same types of
vertices as in \cite{clgr}. In section 5 we prove the renormalizability of
more general acausal theories, where the matter sector contains all
composite operators that have dimensionalities smaller than or equal to
four. We prove the existence of consistent reductions of couplings and the
renormalizability of the models obtained giving an $R$-dependence to the
couplings of power-counting renormalizable theories. In particular, we study
acausal Einstein-Yang-Mills theory with an $R$-dependent gauge coupling.
Section 6 contains the conclusions. In appendices A, B and C we show how to
work out the map $\mathcal{M}$ without using bitensors, write the map $%
\mathcal{M}$ for gravity to the second order in arbitrary spacetime
dimension and perform a detailed analysis of the renormalizability of
Einstein-Yang-Mills theory with an $R$-dependent gauge coupling using the
Batalin-Vilkovisky formalism \cite{bata,wein}.

\section{The map $\mathcal{M}$ and its usage}

\setcounter{equation}{0}

In this section we set the notation and recall the results of ref. \cite
{clgr} that are used or generalized in this paper.

\subsection{Action and field equations for partially quantum, partially
classical theories}

We consider partially quantum, partially classical field theories. Let $%
\varphi _{c}$ denote the classical fields, with action $S_{c}[\varphi _{c}]$%
, and $\varphi $ the quantized fields, with classical action $S[\varphi
,\varphi _{c}]$, embedded in the external $\varphi _{c}$-background. Call $%
\Gamma [\Phi ,\varphi _{c}]$ the generating functional of one-particle
irreducible diagrams obtained quantizing the fields $\varphi $ in the $%
\varphi _{c}$-background, where $\Phi =\left\langle \varphi \right\rangle $.
Then the total action $S_{\text{tot}}[\varphi _{c},\varphi _{q}]$ of the
partially classical, partially quantum theory is defined as 
\begin{equation}
S_{\text{tot}}[\varphi _{c},\varphi _{q}]=S_{c}[\varphi _{c}]+\mathop{\rm Re}%
\Gamma [\Phi ,\varphi _{c}],  \label{stot0}
\end{equation}
where $\varphi _{q}=\Phi $ and $\Phi $ is real if the fields $\varphi $ are
real bosonic, while $\Phi $ is the conjugate of $\overline{\Phi }%
=\left\langle \overline{\varphi }\right\rangle $ if the fields $\varphi $
are complex or fermionic. For example, for classical gravity coupled with
quantum matter, $\varphi _{c}$ is the metric tensor $g_{\mu \nu }$ and $%
S_{c} $ is the Einstein action, so 
\begin{equation}
S_{\text{tot}}[g,\varphi _{q}]=\frac{1}{2\kappa ^{2}}\int \mathrm{d}^{4}x%
\sqrt{-g}\left[ R(g)-2\Lambda \right] +\mathop{\rm Re}\Gamma [\varphi
_{q},g].  \label{stot}
\end{equation}
The field equations are obtained functionally variating the action with
respect to $g_{\mu \nu }$ and $\varphi _{q}$. For example, a simple way to
solve the matter field equations is to set $\varphi _{q}=0$ (or $\varphi
_{q} $ equal to its expectation value, if there is a spontaneous symmetry
breaking). Then the gravitational field equations $\delta S_{\text{tot}%
}[g,0]/\delta g_{\mu \nu }=0$, i.e. 
\begin{equation}
R_{\mu \nu }-\frac{1}{2}g_{\mu \nu }R+g_{\mu \nu }\Lambda =-\kappa ^{2}%
\mathop{\rm Re}\left\langle T_{\mu \nu }\right\rangle ,\qquad \left\langle
T_{\mu \nu }\right\rangle =\frac{2}{\sqrt{-g}}\frac{\delta \Gamma [\varphi
_{q},g]}{\delta g^{\mu \nu }}  \label{moller}
\end{equation}
describe how the spacetime geometry is affected by the quantized matter
fields circulating in the loops.

Another approach to the semi-classical theory, due to Schwinger and Keldysh 
\cite{schwingerkeldysh}, is to replace $\mathop{\rm Re}\left\langle T_{\mu
\nu }\right\rangle $ in (\ref{moller}) with the ``in-in'' expectation value
of the stress tensor, which is both real and causal. Functional methods for
the calculation of in-in expectation values have been developed \cite
{jordan,woodard}. It is important to observe that the renormalization
structure of the theory does not depend on the interpretation of the
right-hand side of the Einstein equations in (\ref{moller}). In particular,
the counterterms $R_{\mu \nu }R^{\mu \nu }$ and $R^{2}$ are identical in the
Schwinger-Kleydish approach \cite{woodard}. The high-energy causality
violations discussed here are an effect due to the renormalization of $%
R_{\mu \nu }R^{\mu \nu }$ and $R^{2}$ by means of field redefinitions of the
metric tensor, so they are independent of the generalization of (\ref{moller}%
) to quantum field theory.

\subsection{The map $\mathcal{M}$}

Consider an action $S$ depending on the fields $\phi $ and modify it into 
\begin{equation}
S^{\prime }[\phi ]=S[\phi ]+S_{i}F_{ij}S_{j},  \label{spri}
\end{equation}
where $F_{ij}$ is symmetric and can contain derivative operators, $%
S_{i}\equiv \delta S/\delta \phi _{i}$ are the $S$-field equations, the
index $i$ stands also for the spacetime point and summation over repeated
indices, including the integration over spacetime points, is understood.
There exists a field redefinition 
\begin{equation}
\phi _{i}^{\prime }=\phi _{i}+\Delta _{ij}S_{j},  \label{redef}
\end{equation}
with $\Delta _{ij}$ symmetric, such that, perturbatively in $F$ and to all
orders in powers of $F$, 
\begin{equation}
S^{\prime }[\phi ]=S[\phi ^{\prime }].  \label{equa}
\end{equation}
Indeed, after a Taylor expansion, it is immediate to see that this equality
is verified if 
\begin{equation}
\Delta _{ij}=F_{ij}-\Delta _{k_{1}i}\Delta _{k_{2}j}\sum_{n=2}^{\infty }%
\frac{1}{n!}S_{k_{1}k_{2}k_{3}\cdots k_{n}}\prod_{l=3}^{n}(\Delta
_{k_{l}m_{l}}S_{m_{l}}).  \label{genfor}
\end{equation}
where $S_{k_{1}\cdots k_{n}}\equiv \delta ^{n}S/(\delta \phi _{k_{1}}\cdots
\delta \phi _{k_{n}})$ and for $n=2$ the product is meant to be unity.
Equation (\ref{genfor}) can be solved recursively for $\Delta $ in powers of 
$F$. The first terms of the solution are 
\begin{eqnarray*}
\Delta _{ij} &=&F_{ij}-\frac{1}{2}F_{ik_{1}}S_{k_{1}k_{2}}F_{k_{2}j}+\frac{1%
}{2}F_{ik_{1}}S_{k_{1}k_{2}}F_{k_{2}k_{3}}S_{k_{3}k_{4}}F_{k_{4}j} \\
&&-\frac{1}{3!}%
F_{ik_{1}}S_{k_{1}k_{2}k_{3}}F_{k_{3}k_{4}}S_{k_{4}}F_{k_{2}j}+\mathcal{O}%
(F^{4}).
\end{eqnarray*}

For example, take an ordinary free field theory 
\[
S[\phi ]=\frac{1}{2}\phi _{i}S_{ij}\phi _{j}. 
\]
Then $S_{k_{1}\cdots k_{n}}=0$ for every $n>2$, while $S_{k_{1}k_{2}}$ is
field-independent and quadratic in the derivatives. The modified action 
\[
S^{\prime }[\phi ]=\frac{1}{2}\phi _{i}\left(
S_{ij}+2S_{ik}F_{km}S_{mj}\right) \phi _{j} 
\]
describes a higher-derivative theory. Equation (\ref{genfor}) simplifies to 
\[
\Delta _{ij}=F_{ij}-\frac{1}{2}\Delta _{k_{1}i}\Delta
_{k_{2}j}S_{k_{1}k_{2}}. 
\]
Its solution reads, in matrix and vector form, 
\[
\Delta =\left( \sqrt{1+2FS}-1\right) S^{-1},\qquad \phi ^{\prime }=\sqrt{%
1+2FS}\phi . 
\]
The map is not just a change of variables, since it changes the degrees of
freedom of the theory.

In the interacting case, we use the free-field limit results to write 
\begin{equation}
\Delta =\left( \sqrt{1+2FS}-1\right) S^{-1}+\mathcal{O}(\phi ),\qquad \phi
^{\prime }=\sqrt{1+2FS}\phi +\mathcal{O}(\phi ^{2}),  \label{cebiso2}
\end{equation}
where $F$ and $S$ are the matrices $F_{ij}$ and $S_{ij}$ calculated at $\phi
=0$. Thus in the acausal theory every $\phi ^{\prime }$-leg gets multiplied
by $1/\sqrt{1+2FS}$.

With a source term the map gives 
\[
S_{\mathrm{HD}}[\phi ,J]\equiv S^{\prime }[\phi ]+\phi _{k}J_{k}=S[\phi
^{\prime }]+\phi _{k}^{\prime }J_{k}^{\prime }(J)\equiv S_{\mathrm{AC}}[\phi
^{\prime },J], 
\]
where 
\begin{equation}
J^{\prime }(J)=\frac{1}{\sqrt{1+2FS}}J.  \label{cebiso}
\end{equation}
The action $S_{\mathrm{HD}}[\phi ,J]$ describes a higher-derivative theory,
while the action $S_{\mathrm{AC}}[\phi ^{\prime },J]$ describes, in general,
an acausal theory. To see this more clearly, take for example $F_{ij}=\alpha
^{2}\delta _{ij}/2$, $S_{ij}=-\Box \delta _{ij}$ then 
\begin{equation}
J^{\prime }(x)=\frac{1}{\sqrt{1-\alpha ^{2}\Box }}J=\int \mathrm{d}%
^{n}x^{\prime }~\mathcal{C}_{n}(x-x^{\prime })J(x^{\prime }),  \label{id2}
\end{equation}
where $n$ is the spacetime dimension and 
\begin{equation}
\mathcal{C}_{n}(x)=\int \frac{\mathrm{d}^{n}p}{(2\pi )^{n}}\frac{\mathrm{e}%
^{-ip\cdot x}}{\sqrt{1+\alpha ^{2}p^{2}}}.  \label{ava}
\end{equation}

The Fourier transform (\ref{ava}) has to be defined with an appropriate
prescription. The degrees of freedom that are responsible for the
instabilities in the higher-derivative model are suppressed demanding that
the prescription be regular in the limit $\alpha \rightarrow 0$. For
example, 
\begin{equation}
\mathcal{C}_{n}^{\text{F}}(x)=\int \frac{\mathrm{d}^{n}p}{(2\pi )^{n}}\frac{%
\mathrm{e}^{-ip\cdot x}}{\sqrt{1+\alpha ^{2}p^{2}+i\varepsilon }}.
\label{cnf}
\end{equation}
Observe that $\mathcal{C}_{n}^{\text{F}}(x)$ is complex, but the definition (%
\ref{stot0}) of the action takes care of this.

Although the map is perturbative in $F$, formula (\ref{cnf}) allows us to
study some effects of the resummation of derivatives. The function $\mathcal{%
C}_{n}^{\text{F}}(x)$ does not vanish outside the past light cone, so
causality is violated. In some cases causal prescriptions for $\mathcal{C}%
_{n}(x)$ exist, but when the radiative corrections are taken into account, $%
\alpha ^{2}$ runs and its logarithmic dependence in general spoils the
causal prescriptions. The violation of causality is the price paid for the
elimination of instabilities.

The function $\mathcal{C}_{n}^{\text{F}}(x)$ tends to zero or rapidly
oscillates for $|x^{2}|\gg |\alpha ^{2}|$, so the causality violations can
be experimentally tested only at distances of the order of 
\begin{equation}
\Delta x\sim |\alpha |  \label{range}
\end{equation}
and become physically unobservable at distances much larger than this bound.

The functional derivatives $S_{ijk\cdots }$ are bi-, tri-tensor densities,
etc., and involve several Dirac delta functions. It can be cumbersome to
preserve manifest general covariance using these objects. Fortunately, it is
not really necessary to work with them, because they appear only in the
intermediate formulas. In the appendix we show how to work without them,
using only tensors and tensor densities.

Another approach to remove instabilities is known in the literature as the 
\textit{regular reduction} of the order of the differential equation. It
mimics a manipulation usually learnt in connection with the Abraham-Lorentz
force in classical electrodynamics \cite{jackson} and extends it to the case
of gravity \cite{bel,parkersimon}. The regular reduction is not a field
redefinition, but a manipulation of the field equations, which leaves the
metric tensor unchanged. The map $\mathcal{M}$, on the other hand, is
designed to work efficiently in combination with renormalization.

\subsection{Usage of the map $\mathcal{M}$}

The map $\mathcal{M}$ can be used to convert a higher-derivative theory of
classical gravity coupled with quantum matter into an acausal theory,
preserving the renormalizability. Consider the higher-derivative theory

\begin{equation}
S_{\text{HD}}[\overline{g},\varphi ,\lambda ,a,b,\kappa ]=\frac{1}{2\kappa
^{2}}\int \mathrm{d}^{4}x\sqrt{-\overline{g}}\left( \overline{R}+a\overline{R%
}_{\mu \nu }\overline{R}^{\mu \nu }+b\overline{R}^{2}\right) +S_{m}[\varphi ,%
\overline{g},\lambda ],  \label{hg}
\end{equation}
where $\overline{R}_{\mu \nu }$, $\overline{R}$ are the Ricci tensor and
scalar curvature of the metric $\overline{g}$. Here $S_{m}$ is the
power-counting renormalizable matter action embedded in curved background.
For simplicity assume that $S_{m}$ does not contain masses and
super-renormalizable parameters and use the dimensional-regularization
technique. Then no cosmological constant is generated by renormalization.
The arguments below can be generalized straightforwardly to include these
parameters, together with the cosmological constant.

Obviously $S_{\text{HD}}$ is renormalizable, but physically unsatisfactory
due to the higher-derivative kinetic terms in the gravitational sector.
However, the theorem just proved ensures that there exists a map $\overline{g%
}=\overline{G}(g,a,b)$ such that 
\[
\int \mathrm{d}^{4}x\sqrt{-\overline{g}}\left[ \overline{R}(\overline{g})+a%
\overline{R}_{\mu \nu }\overline{R}^{\mu \nu }(\overline{g})+b\overline{R}%
^{2}(\overline{g})\right] =\int \mathrm{d}^{4}x\sqrt{-g}R(g) 
\]
and reabsorbs the higher-derivative gravitational kinetic terms into the
Einstein term. Applied to (\ref{hg}), the map generates new vertices that
couple matter with gravity, and defines a new action $S_{\text{AC}}$ such
that 
\begin{equation}
S_{\text{HD}}[\overline{G}(g,a,b),\varphi ,\lambda ,a,b,\kappa ]=S_{\text{AC}%
}[g,\varphi ,\lambda ,a,b,\kappa ].  \label{ide}
\end{equation}
The action $S_{\text{AC}}$ has the form 
\begin{equation}
S_{\text{AC}}[g,\varphi ,\lambda ,\lambda ^{\prime },\kappa ]=\frac{1}{%
2\kappa ^{2}}\int \mathrm{d}^{4}x\sqrt{-g}R+S_{m}[\varphi ,g,\lambda
]+\Delta S_{m}[\varphi ,g,\lambda ,\lambda ^{\prime }],  \label{acca2}
\end{equation}
where 
\begin{equation}
\Delta S_{m}=\int \mathrm{d}^{4}x\sqrt{-g}\ \left[ -\frac{a}{2}T_{m}^{\mu
\nu }R_{\mu \nu }+\frac{1}{4}(a+2b)RT_{m}\right] +\mathcal{O}%
(a^{2},b^{2},ab),  \label{avert}
\end{equation}
where $T_{m}^{\mu \nu }=-(2/\sqrt{-g})(\delta S_{m}/\delta g_{\mu \nu })$ is
the stress-tensor of the uncorrected matter sector and $T_{m}$ denotes its
trace. Precisely, $\Delta S_{m}$ does not contain any kinetic contributions
and is made of vertices that are either proportional to $T_{m}^{\mu \nu }$
and (covariant derivatives of) the Ricci tensor, or quadratically
proportional to (covariant derivatives of) the Ricci tensor. More details
are given in the appendix.

Formula (\ref{ide}) is the relation between the classical actions. Analogous
relations hold for the bare and renormalized actions, when the matter fields
are quantized, and the generating functionals $\Gamma $: 
\begin{equation}
\Gamma _{\text{AC}}[g,\Phi ,\lambda ,a,b,\kappa ]=\Gamma _{\text{HD}}[%
\overline{G}(g,a,b),\Phi ,\lambda ,a,b,\kappa ].  \label{gamma}
\end{equation}
The total actions $S_{\text{tot AC}}$ and $S_{\text{tot HD}}$ of (\ref{stot0}%
) follow from their definitions. Observe that the resummed map $\overline{g}=%
\overline{G}(g,a,b)$ is complex, in general, due to the prescription (\ref
{cnf}). The acausal action $S_{\text{tot AC}}$ is defined taking the real
part of $\Gamma _{\text{AC}}$ with $g$ and $\Phi $ real, \textit{after}
applying the map $\mathcal{M}$ to $\Gamma _{\text{HD}}$, which is not the
same as applying the map $\mathcal{M}$ to $S_{\text{tot HD}}$.

The map $\mathcal{M}$ preserves the renormalizability of the theory. Indeed,
the function $\overline{g}=\overline{G}(g,a,b)$ is finite and does not
depend on the quantum fields, so $\Gamma _{\text{AC}}$ is convergent because 
$\Gamma _{\text{HD}}$ is.

Finally, the map is not just a change of variables, but changes the physics,
since it eliminates the unwanted degrees of freedom at the price of
introducing violations of causality at small distances. In the expansion
around flat space, $g_{\mu \nu }=\eta _{\mu \nu }+2\kappa \phi _{\mu \nu }$, 
$\eta _{\mu \nu }=$diag$(1,-1,-,1-1)$, the traceless part $\widetilde{\phi }%
_{\mu \nu }$ of $\phi _{\mu \nu }$ is mapped as \cite{clgr} 
\begin{equation}
\widetilde{\phi }_{\mu \nu }=\frac{1}{\sqrt{1-a\Box }}\widetilde{\phi }_{\mu
\nu }^{\prime }.  \label{traceless}
\end{equation}
The other components of $\phi _{\mu \nu }$ are multiplied by $1/\sqrt{%
1-a\Box }$ or $\sqrt{1-b^{\prime }\Box }$, where $b^{\prime }=-2(a+3b)$ (see 
\cite{clgr} for other explicit formulas). Thus, the causality violations due
to the map $\mathcal{M}$ become detectable at distances of the order of $%
\sqrt{|a|}$, $\sqrt{|b^{\prime }|}$ or smaller.

\section{Acausal Einstein-Yang-Mills theory}

\setcounter{equation}{0}

For definiteness, in this section and in other sections of the paper, we
consider an explicit model, non-Abelian Yang-Mills theory coupled with
classical gravity. However, most of the properties that we uncover are
valid, with minor modifications, in every power-counting renormalizable
theory coupled with classical gravity. We recall a number of results from
the literature \cite{hathrell,hathrellQED,freeman} that will be useful in
section 5, where we let the gauge coupling depend on the scalar curvature $R$
of spacetime. We apply the map $\mathcal{M}$ and describe the acausal
Einstein-Yang-Mills theory in detail, in particular its renormalization. The
metric tensor has a non-trivial running although it is not quantized. The
runnings of $a$ and $b^{\prime }$ are related with the coefficients of the
trace anomaly of Yang-Mills theory in external gravity and can be studied
exactly in a large class of models, those that interpolate between UV\ and
IR\ conformal fixed points.

The lagrangian of the higher-derivative renormalizable theory is 
\begin{equation}
\frac{\mathcal{L}_{\mathrm{HD}}}{\sqrt{-\overline{g}}}=\frac{\overline{R}}{%
2\kappa ^{2}}+\xi \overline{W}^{2}+\zeta \overline{G}_{B}+\frac{\eta }{%
(n-1)^{2}}\overline{R}^{2}-\frac{1}{4\alpha }F_{\mu \nu }^{a}F^{a\hspace{%
0.01in}\mu \nu }~,  \label{basis}
\end{equation}
where $F_{\mu \nu }^{a}$ is the Yang-Mills field strength, $\alpha $ is the
squared gauge coupling, indices are raised and lowered with the metric $%
\overline{g}_{\mu \nu }$, $\overline{W}^{2}$ is the square of the Weyl
tensor and $\overline{G}_{B}$ is the Gauss-Bonnet density, 
\begin{equation}
\overline{W}^{2}=\overline{R}_{\mu \nu \rho \sigma }\overline{R}^{\mu \nu
\rho \sigma }-\frac{4}{n-2}\overline{R}_{\mu \nu }\overline{R}^{\mu \nu }+%
\frac{2}{(n-1)(n-2)}\overline{R}^{2},\qquad \overline{G}_{B}=\overline{R}%
_{\mu \nu \rho \sigma }\overline{R}^{\mu \nu \rho \sigma }-4\overline{R}%
_{\mu \nu }\overline{R}^{\mu \nu }+\overline{R}^{2}.
\end{equation}

We use the dimensional-regularization technique, $n=4-\varepsilon $ being
the continued spacetime dimension, and the minimal subtraction scheme. Since 
$\sqrt{-g}\overline{G}_{B}$ is a total derivative in four dimensions,
assuming that the metric tends to flat space at infinity with an appropriate
velocity, the integral 
\begin{equation}
\int \mathrm{d}^{n}x\sqrt{-g}\ \overline{G}_{B}  \label{int}
\end{equation}
is ``evanescent'' \cite{collins}. This means that a counterterm proportional
to (\ref{int}), for example $1/\varepsilon $ times (\ref{int}), is not a
true divergence, but amounts to a finite local correction to the quantum
action, that is to say a scheme redefinition. Thus the parameter $\zeta $
does not affect the physical quantities and should not be considered as a
new physical coupling. Moreover, keeping $n$ arbitrary and expanding the
metric around flat space, the integral (\ref{int}) does not contain kinetic
contributions \cite{abse}. Writing $\overline{W}^{2}$ as a linear
combination of $\overline{G}_{B}$, $\overline{R}_{\mu \nu }\overline{R}^{\mu
\nu }$ and $\overline{R}^{2}$, the relations with the couplings $a$ and $b$
used in the previous section read 
\begin{equation}
\frac{a}{2\kappa ^{2}}=\frac{4(n-3)}{n-2}\xi ,\qquad \frac{b}{2\kappa ^{2}}=%
\frac{\eta }{(n-1)^{2}}-\frac{n(n-3)}{(n-1)(n-2)}\xi .  \label{copli}
\end{equation}

Renormalization in curved background is achieved by means of the usual
renormalization constants, plus additional renormalization constants for $%
\xi $, $\zeta $ and $\eta $. The metric $\overline{g}_{\mu \nu }$ and the
Newton constant $\kappa $ are not renormalized. In the minimal subtraction
scheme, $\alpha $ has beta function 
\[
\mu \frac{\mathrm{d}\alpha }{\mathrm{d}\mu }=\widehat{\beta }(\alpha
)=-\varepsilon \alpha +\beta (\alpha ),\qquad \beta (\alpha )=-\frac{22}{3}%
\frac{\alpha ^{2}C(G)}{(4\pi )^{2}}+\mathcal{O}(\alpha ^{3}). 
\]
The bare coupling $\xi _{\mathrm{B}}$ is related with the renormalized
coupling $\xi $ by the formula 
\begin{equation}
\xi _{\mathrm{B}}=\mu ^{-\varepsilon }\left( \xi +L_{\xi }\right) ,\qquad
L_{\xi }=\sum_{i=1}^{\infty }\frac{\xi _{i}(\alpha )}{\varepsilon ^{i}}.
\label{ab}
\end{equation}
Standard RG relations implied with the finiteness of the beta function $%
\beta _{\xi }$ give, in the minimal subtraction scheme, 
\begin{equation}
\mu \frac{\mathrm{d}\xi }{\mathrm{d}\mu }=\widehat{\beta }_{\xi
}=\varepsilon \xi +\beta _{\xi },\qquad \beta _{\xi }=\varepsilon L_{\xi
}-\mu \frac{\mathrm{d}L_{\xi }}{\mathrm{d}\mu }=\frac{\mathrm{d}}{\mathrm{d}%
\alpha }\left( \alpha \xi _{1}(\alpha )\right)  \label{abeta}
\end{equation}
and recursively relate the functions $\xi _{i}(\alpha )$ by 
\begin{equation}
\frac{\mathrm{d}}{\mathrm{d}\alpha }\left( \alpha \xi _{k+1}(\alpha )\right)
=\beta (\alpha )\frac{\mathrm{d}\xi _{k}(\alpha )}{\mathrm{d}\alpha }.
\label{ab2}
\end{equation}
Formulas similar to (\ref{ab}), (\ref{ab2}) and (\ref{abeta}) hold for the
couplings $\zeta $ and $\eta $ and define beta functions $\beta _{\zeta }$, $%
\beta _{\eta }$, which are, as well as $\beta _{\xi }$, just functions of $%
\alpha $. The lowest-order results \cite{freeman} are 
\begin{eqnarray}
\beta _{\xi } &=&\frac{\dim G}{(4\pi )^{2}}\left( -\frac{1}{10}+\frac{2}{9}%
\frac{\alpha C(G)}{(4\pi )^{2}}\right) +\mathcal{O}(\alpha ^{2}),  \nonumber
\\
\beta _{\zeta } &=&-\frac{\dim G}{(4\pi )^{2}}\left( \frac{31}{180}+\frac{17%
}{12}\frac{\alpha ^{2}C^{2}(G)}{(4\pi )^{4}}\right) +\mathcal{O}(\alpha
^{3}),  \label{trick} \\
\beta _{\eta } &=&\frac{\dim G}{(4\pi )^{2}}\frac{187}{54}\frac{\alpha
^{3}C^{3}(G)}{(4\pi )^{6}}+\mathcal{O}(\alpha ^{4}).  \nonumber
\end{eqnarray}

\bigskip

Formulas for the beta functions of $a$ and $b$ can be written using the
relations (\ref{ab}) and (\ref{abeta}), e.g. 
\[
\beta _{a}=\frac{8(n-3)}{n-2}\kappa ^{2}\beta _{\xi }(\alpha ), 
\]
and are valid in every power-counting renormalizable theory coupled with
classical gravity.

In flat space there are many examples of quantum field theories whose
renormalization-group flow interpolates between ultraviolet and infrared
conformal field theories. Conformal field theories are characterized, among
the other things, by certain quantities, called ``central charges'' (for
definitions and properties, see for example \cite{noi}). Popular models
interpolating between UV\ and IR\ conformal fixed points are provided by
supersymmetric gauge theories in the ``conformal window'', where the values
of the central charges can be calculated exactly also at the interacting
fixed points \cite{noi}. Massless QCD can be considered a model with
analogous features, because it interpolates between a free UV theory of
quarks and gluons and a free IR theory of massless pions. When these models
are coupled with classical gravity, the running of $a$ can be studied
throughout the RG\ flow. Indeed, the function $\beta _{\xi }(\alpha )$ is
well-behaved and interpolates between (minus) the UV and IR\ values of the
central charge $c$, 
\[
\lim_{\mathrm{UV}\text{ }\mathrm{(IR)}}\beta _{\xi }(\alpha )=-c_{\mathrm{UV}%
\text{ }\mathrm{(IR)}}. 
\]
In massless free field theories the quantity $c$ is equal to 
\[
c=\frac{12n_{v}+6n_{f}+n_{s}}{120(4\pi )^{2}}, 
\]
where $n_{v,f,s}$ are the numbers of vectors, fermions and scalars. At a
conformal fixed point the $a$-running is just 
\[
a(-p^{2})=\overline{a}-2c\kappa ^{2}\ln \frac{-p^{2}}{\mu ^{2}}, 
\]
where $\overline{a}=a(\mu ^{2})$, so (\ref{ava}) (with $\alpha
^{2}\rightarrow a$) is renormalization-group improved to 
\[
\mathcal{C}_{n}(x)=\int \frac{\mathrm{d}^{n}p}{(2\pi )^{n}}\frac{\mathrm{e}%
^{-ip\cdot x}}{\sqrt{1+a(-p^{2})p^{2}}}. 
\]
Since the radiative corrections to $a$ are just logarithmic, the
large-distance behavior of $\mathcal{C}_{n}(x)$ is still causal. The
violations of causality can be appreciated at energies $E$ such that 
\[
a(E^{2})E^{2}\sim 1. 
\]

\bigskip

Now we apply the map $\mathcal{M}$ to the higher-derivative model (\ref
{basis}) and describe the acausal theory that we obtain. For the reasons
explained above, we can keep $n=4$ without loosing information. From
appendix B, we have 
\begin{eqnarray}
\overline{g}_{\mu \nu } &=&g_{\mu \nu }+aR_{\mu \nu }-\frac{a+2b}{2}g_{\mu
\nu }R+\frac{3a^{2}}{4}\Box R_{\mu \nu }-\frac{3a(a+2b)}{4}\nabla _{\mu
}\nabla _{\nu }R-abRR_{\mu \nu }+\frac{1}{2}a^{2}R_{\mu }^{\lambda
}R_{\lambda \nu }  \nonumber \\
&&-\frac{3}{2}a^{2}R_{\mu \alpha \nu \beta }R^{\alpha \beta }+\frac{1}{8}
g_{\mu \nu }\left\{ 3(a+2b)\left( a+6b\right) \Box R+2a\left( 3a+4b\right)
R_{\alpha \beta }R^{\alpha \beta }-a^{2}R^{2}\right\}  \nonumber \\
&&+\mathcal{O}(a^{3},a^{2}b,ab^{2}).  \label{gHD}
\end{eqnarray}

The acausal Einstein-Yang-Mills lagrangian has the form 
\begin{equation}
\mathcal{L}_{\mathrm{EYM}\text{-}\mathrm{AC}}=\frac{1}{2\kappa ^{2}}\sqrt{-g}%
R-\frac{1}{4\alpha }\sqrt{-g}\left\{ F_{\mu \nu }^{a}F^{a\hspace{0.01in}\mu
\nu }~H(g)+T_{\mu \nu }K^{\mu \nu }(g)+\Upsilon _{\mu \nu \rho \sigma
}L^{\mu \nu \rho \sigma }(g)\right\} ,  \label{lqed}
\end{equation}
where $T_{\mu \nu }$ is the unperturbed stress tensor and $\Upsilon _{\mu
\nu \rho \sigma }$ is the traceless operator $F_{\mu \nu }F_{\rho \sigma }$, 
\begin{eqnarray}
T_{\mu \nu } &=&-F_{\mu \alpha }^{a}F_{\nu }^{a\hspace{0.03in}\alpha }+\frac{%
1}{4}g_{\mu \nu }F^{2},\qquad \qquad \qquad  \label{tmn} \\
\hspace{-0.1in}\Upsilon _{\mu \nu \rho \sigma } &=&F_{\mu \nu }^{a}F_{\rho
\sigma }^{a}+\frac{1}{2}\left( g_{\mu \rho }T_{\nu \sigma }-g_{\mu \sigma
}T_{\nu \rho }-g_{\nu \rho }T_{\mu \sigma }+g_{\nu \sigma }T_{\mu \rho
}\right) -\frac{1}{12}\left( g_{\mu \rho }g_{\nu \sigma }-g_{\mu \sigma
}g_{\nu \rho }\right) F^{2}.  \label{upsilon}
\end{eqnarray}
where $F^{2}\equiv F_{\alpha \beta }^{a}F^{a\hspace{0.01in}\alpha \beta }$.
The coefficients $H(g)$, $K^{\mu \nu }(g)$ and $L^{\mu \nu \rho \sigma }(g)$
are non-polynomial tensorial functions of the metric tensor. Moreover, $%
K^{\mu \nu }(g)$ is proportional to the Ricci tensor or its covariant
derivatives, while $L^{\mu \nu \rho \sigma }(g)$ and $H(g)-1$ are squarely
proportional to the Ricci tensor or its covariant derivatives.

Thus, the acausal Einstein-Yang-Mills theory is just Yang-Mills theory with
two composite operators besides $F^{2}$, which are coupled with suitable
metric-dependent external sources. To the second order, using the results of
appendix B we find 
\begin{eqnarray}
&&H(g)=1+\frac{1}{6}a^{2}R_{\alpha \beta }R^{\alpha \beta }-\frac{1}{24}%
a^{2}R^{2},\qquad L^{\mu \nu \rho \sigma }=a^{2}R^{\mu \rho }R^{\nu \sigma },
\label{source} \\
&&K^{\mu \nu }(g)=2aR^{\mu \nu }+\frac{3}{2}a^{2}\Box R^{\mu \nu
}+a^{2}RR^{\mu \nu }-3a^{2}R^{\mu \alpha \nu \beta }R_{\alpha \beta }-\frac{%
3a(a+2b)}{2}\nabla ^{\mu }\nabla ^{\nu }R.  \label{sources2}
\end{eqnarray}

The renormalization of the acausal theory works as follows: the couplings $a$
and $b$ renormalize according to the formulas (\ref{copli}) and (\ref{trick}%
). Moreover, the metric tensor $g_{\mu \nu }$, although it is not quantized,
has a non-trivial running, induced by the $a$- and $b$- runnings, which can
be determined using the non-renormalization of $\overline{g}_{\mu \nu }$ in (%
\ref{gHD}). Thus, 
\[
\mu \frac{\mathrm{d}\overline{g}_{\rho \sigma }}{\mathrm{d}\mu }=0\qquad
\Longrightarrow \qquad \mu \frac{\mathrm{d}g_{\rho \sigma }}{\mathrm{d}\mu }%
=-\beta _{a}R_{\rho \sigma }+\frac{1}{2}(\beta _{a}+2\beta _{b})g_{\mu \nu
}R+\mathcal{O}(a\kappa ^{2},b\kappa ^{2}).
\]
The Gauss-Bonnet term is sent into itself by the map $\mathcal{M}$. The $n$%
-dependence of the map $\mathcal{M}$ away from $n=4$ generates a number of
new evanescent terms in the acausal theory (see appendix B).

\section{First generalization of renormalizable theories}

\setcounter{equation}{0}

In the rest of the paper we study more general renormalizable theories of
classical gravity coupled with quantized fields. The divergences are
subtracted with a finite or reduced set of independent couplings. The
acausal models can still be obtained from higher-derivative models using the
map $\mathcal{M}$. We generalize the constructions of the previous sections
in two main ways. In the present section we focus on theories whose
non-renormalizable perturbation has a head of the form (\ref{avert}), namely
proportional to the Ricci tensor and the energy-momentum tensor of the
unperturbed matter sector. In section 5 we study non-renormalizable
deformations whose heads contain more general matter operators, let the
couplings depend on the scalar curvature of spacetime, and so on.

We consider theories of the form 
\begin{equation}
S_{\text{AC}}=\frac{1}{2\kappa ^{2}}\int \mathrm{d}^{4}x\sqrt{-g}%
R+S_{m}(\varphi ,g,\lambda )+\Delta S_{m}(\varphi ,g,\lambda ^{\prime }).
\label{acca3}
\end{equation}
Here $S_{m}(\varphi ,g,\lambda )$ denotes the power-counting renormalizable
matter action embedded in external gravity, with couplings $\lambda $. It
includes the non-minimal terms of dimensionality equal to four. For the
moment, we set the masses and other super-renormalizable parameters to zero.
Instead, $\Delta S_{m}(\varphi ,g,\lambda ^{\prime })$ denotes the vertices
of dimensionalities greater than four, multiplied by couplings $\lambda
^{\prime }$. Neither $S_{m}$ nor $\Delta S_{m}$ include purely gravitational
terms.

Assume that $\Delta S_{m}$ contains only a finite number of matter
operators, all those that have dimensionality four or smaller than four, are
gauge invariant, covariant under diffeomorphisms, invariant under the global
symmetries of the theory, not necessarily scalar, and can be contracted with
tensors constructed with the metric in a non-trivial way. For example, in
the case of Yang-Mills theory, the operators are just $F^{2}$, plus $T^{\mu
\nu }$ (from now on called $T_{m}^{\mu \nu }$) and $\Upsilon ^{\mu \nu \rho
\sigma }$, formulas (\ref{tmn}) and (\ref{upsilon}). They can be contracted
with a variety of tensors constructed with the metric, giving e.g. 
\begin{equation}
RF^{2},\qquad R_{\mu \nu }T_{m}^{\mu \nu },\qquad R_{\mu \nu \rho \sigma
}\Upsilon ^{\mu \nu \rho \sigma },\qquad R_{\mu \rho }R_{\nu \sigma
}\Upsilon ^{\mu \nu \rho \sigma },\qquad \nabla _{\nu }\nabla _{\sigma
}R_{\mu \rho }\Upsilon ^{\mu \nu \rho \sigma },  \label{aneipol}
\end{equation}
etc. Moreover, assume that $\Delta S_{m}$ is proportional to the Ricci
tensor or its covariant derivatives and the vertices that depend linearly on
the Ricci tensor are also proportional to the stress tensor $T_{m}^{\mu \nu
} $ of the unperturbed action. Collecting these assumptions into explicit
formulas, $\Delta S_{m}$ has the structure 
\begin{eqnarray}
\Delta S_{m} &=&\Delta _{1}S_{m}+\Delta _{2}S_{m},\qquad \Delta
_{1}S_{m}=\int \mathrm{d}^{4}x\sqrt{-g}f_{\mu \nu }(g_{\rho \sigma
})T_{m}^{\mu \nu },  \label{first} \\
\qquad \Delta _{2}S_{m} &=&\mathcal{O}(R_{\mu \nu }^{2},\varphi ),\qquad
\qquad T_{m}^{\mu \nu }=-\frac{2}{\sqrt{-g}}\frac{\delta S_{m}}{\delta
g_{\mu \nu }(x)},  \nonumber
\end{eqnarray}
\noindent where $f_{\mu \nu }(g_{\rho \sigma })$ is a tensor depending on
the metric, proportional to the Ricci tensor (or its covariant derivatives),
of the form 
\[
f_{\mu \nu }(g_{\rho \sigma })=\lambda _{1}^{\prime }R_{\mu \nu }+\lambda
_{2}^{\prime }Rg_{\mu \nu }+\mathcal{O}(\nabla ^{2}R_{\rho \sigma }). 
\]
and $\Delta _{2}S_{m}$ contains the prescribed set of matter operators. Thus
for example, the second term of (\ref{aneipol}) belongs to $\Delta _{1}S_{m}$
while the forth belongs to $\Delta _{2}S_{m}$. Instead, the first, third and
fifth term of (\ref{aneipol}) do not belong to $\Delta S_{m}$, unless they
are multiplied by at least 1, 2 and 1 extra powers of the Ricci tensor,
respectively.

A simple example is a generalization of acausal Einstein-Yang-Mills theory,
described by a lagrangian of the form (\ref{lqed}), where however $H(g)$, $%
K^{\mu \nu }(g)$ and $L^{\mu \nu \rho \sigma }(g)$ do not have necessarily
the expressions (\ref{source}) and (\ref{sources2}), inherited applying the
map $\mathcal{M}$ to (\ref{basis}), but are arbitrary functions of the
metric, subject only to the restrictions that $K^{\mu \nu }(g)$ be
proportional to the Ricci tensor or its covariant derivatives, and $L^{\mu
\nu \rho \sigma }(g)$ and $H(g)-1$ be squarely proportional to the Ricci
tensor or its covariant derivatives.

\bigskip

We wish to prove that the action (\ref{acca3}) is renormalizable in the form
(\ref{acca3}). We first prove that this result follows from the
renormalizability of the theory 
\begin{equation}
S_{\mathrm{HD}}=\frac{1}{2\kappa ^{2}}\int \mathrm{d}^{4}x\sqrt{-g}R+\int 
\mathrm{d}^{4}x\sqrt{-g}R_{\mu \nu }\mathcal{T}^{\mu \nu \rho \sigma
}(\lambda ^{\prime })R_{\rho \sigma }+S_{m}(\varphi ,g,\lambda )+\Delta
S_{m}(\varphi ,g,\lambda ,\lambda ^{\prime }).  \label{saccadi}
\end{equation}
Here $\Delta S_{m}$ is not the same as in (\ref{acca3}), but it is subject
to the same restrictions as the $\Delta S_{m}$ of (\ref{acca3}). We use the
same notation, since no confusion can arise, because what is important here
is just the structure of $\Delta S_{m}$, not its precise value. The tensor $%
\mathcal{T}^{\mu \nu \rho \sigma }$ is a (possibly differential) operator
that depends only on the metric and can contain covariant derivatives acting
on the Ricci tensors to its left and to its right. It can contain other
couplings $\lambda ^{\prime }$.

Later we prove the renormalizability of (\ref{saccadi}). First we prove that
the renormalizability of (\ref{saccadi}) implies the renormalizability of (%
\ref{acca3}), applying the map $\mathcal{M}$.

The theorem of section 2 ensures that there exists a map $g_{\mu \nu
}^{\prime }=g_{\mu \nu }^{\prime }(g)=g_{\mu \nu }+\Delta g_{\mu \nu }(g)$,
with $\Delta g_{\mu \nu }$ proportional to the Ricci tensor and its
covariant derivatives, that reabsorbs the pure gravitational part of (\ref
{saccadi}) inside the Einstein term, namely such that 
\begin{equation}
\frac{1}{2\kappa ^{2}}\int \mathrm{d}^{4}x\sqrt{-g}R+\int \mathrm{d}^{4}x%
\sqrt{-g}R_{\mu \nu }\mathcal{T}^{\mu \nu \rho \sigma }R_{\rho \sigma }=%
\frac{1}{2\kappa ^{2}}\int \mathrm{d}^{4}x\sqrt{-g^{\prime }}R(g^{\prime }).
\label{aida}
\end{equation}
On the other hand, acting on $S_{m}+\Delta S_{m}$ the redefinition produces
terms that fall again in the classes $S_{m}$, $\Delta _{1}S_{m}$ and $\Delta
_{2}S_{m}$, as we now prove.

First, the matter operators generated by this process can have at most
dimensionality four, since $S_{m}$ and $\Delta S_{m}$ contain only, and all,
such operators. Second, to the second and higher orders in $\Delta g_{\mu
\nu }$ the terms generated by the field redefinition are certainly quadratic
in the Ricci tensor or its covariant derivatives, so they fall in the class $%
\Delta _{2}S_{m}$. Finally, to the first order in $\Delta g_{\mu \nu }$,

\noindent $i$) the terms generated varying $S_{m}$ are certainly
proportional to the energy-momentum tensor $T_{m}^{\mu \nu }$ of the
unperturbed action $S_{m}$ and proportional to the Ricci tensor, so they
fall in the class $\Delta _{1}S_{m}$;

\noindent $ii$) the terms generated varying $\Delta _{1}S_{m}$ can be of two
types:

$a$) those obtained varying the metric outside $T_{m}^{\mu \nu }$ are still
proportional to $T_{m}^{\mu \nu }$ and at least linearly proportional to the
Ricci tensor, so they fall either in the class $\Delta _{1}S_{m}$, or in the
class $\Delta _{2}S_{m}$;

$b$) those obtained varying the metric inside $T_{m}^{\mu \nu }$ are not
necessarily proportional to $T_{m}^{\mu \nu }$, but they are at least
squarely proportional to the Ricci tensor, so they fall in the class $\Delta
_{2}S_{m}$;

\noindent $iii$) the terms generated varying $\Delta _{2}S_{m}$ fall
necessarily in the class $\Delta _{2}S_{m}$.

\bigskip

Thus, it is sufficient to prove the renormalizability of (\ref{saccadi}). We
first describe how the renormalization of the theory (\ref{hg}) works,
namely the particular case $\Delta S_{m}=0$. We write 
\begin{equation}
S_{\mathrm{HD}}^{(0)}(\varphi ,g,\lambda ,a,b,\kappa )=\frac{1}{2\kappa ^{2}}%
\int \mathrm{d}^{4}x\sqrt{-g}\left( R+aR_{\mu \nu }R^{\mu \nu
}+bR^{2}\right) +S_{m}(\varphi ,g,\lambda )  \label{sHD0}
\end{equation}
and denote the quantities relative to this theory with the subscript $0$.
The counterterms generated by $S_{\mathrm{HD}}^{(0)}$ are of the following
types:

\noindent 1) counterterms proportional to the vertices of $S_{m}(\varphi
,g,\lambda )$. They are subtracted multiplying the matter fields and the
couplings $\lambda $ contained in $S_{m}$ by appropriate renormalization
constants. The wave-function renormalization constants coincide with the
flat-space ones. The renormalization constant of a vertex that survives the
flat-space limit coincides with the flat-space one. If the vertex vanishes
in the flat-space limit, as the non-minimal term $R\varphi ^{2}$ for scalar
fields $\varphi $, its renormalization constant has to be calculated anew.

\noindent 2) BRST-exact terms, not included in (\ref{sHD0}), which do not
affect the physical sector of the theory;

\noindent 3) purely gravitational counterterms, proportional to 
\begin{equation}
R_{\mu \nu }R^{\mu \nu },\qquad R^{2}.  \label{countergrav}
\end{equation}
As usual, the Gauss-Bonnet identity is used to convert $R_{\mu \nu \rho
\sigma }R^{\mu \nu \rho \sigma }$ into a linear combination of (\ref
{countergrav}). The counterterms (\ref{countergrav}) obviously fall in the
pure gravitational sector of (\ref{saccadi}).

Proceeding inductively, denote with $\Gamma _{0\hspace{0.01in}}^{(n)}$ the
generating functional of one-particle irreducible diagrams of the theory $S_{%
\mathrm{HD}}^{(0)}$ renormalized up to $n$ loops. Then its $(n+1)$-loop
divergent part $\Gamma _{0\hspace{0.02in}\text{div}}^{(n)}$ is local and has
the form

\begin{equation}
\Gamma _{0\hspace{0.02in}\text{div}}^{(n)}=\frac{\partial S_{m}}{\partial
\lambda }\Delta _{n}\lambda +\varphi \frac{\delta S_{m}}{\delta \varphi }%
\Delta _{n}Z_{\varphi }+\frac{1}{2\kappa ^{2}}\int \mathrm{d}^{4}x\sqrt{-g}%
\left( \Delta _{n}a\hspace{0.02in}R_{\mu \nu }R^{\mu \nu }+\Delta _{n}b%
\hspace{0.02in}R^{2}\right) ,  \label{account}
\end{equation}
plus BRST-exact terms, where $\Delta _{n}\lambda $, $\Delta _{n}Z_{\varphi }$%
, $\Delta _{n}a$ and $\Delta _{n}b$ are appropriate divergent constants. The
divergences (\ref{account}) are subtracted by the redefinitions 
\begin{equation}
\varphi \rightarrow \varphi -\varphi \Delta _{n}Z_{\varphi },\qquad \lambda
\rightarrow \lambda -\Delta _{n}\lambda ,\qquad a\rightarrow a-\Delta
_{n}a,\qquad b\rightarrow b-\Delta _{n}b,  \label{redefa}
\end{equation}

\noindent and the counterterms are collected in 
\begin{equation}
S_{\mathrm{HD}}^{(0)}\left( \varphi -\varphi \Delta _{n}Z_{\varphi
},g,\lambda -\Delta _{n}\lambda ,a-\Delta _{n}a,b-\Delta _{n}b,\kappa
\right) .  \label{verysame}
\end{equation}

\bigskip

Now consider the theory (\ref{saccadi}). We prove that its divergences are
renormalized with the same redefinitions (\ref{redefa}), plus redefinitions
of the parameters contained in $\Delta S_{m}$. Consider first the diagrams
that do not contain insertions of $\Delta S_{m}$. Their divergent parts have
to be reabsorbed with the redefinitions (\ref{redefa}), but now those
redefinitions have to be performed on the full action $S_{\mathrm{HD}}$, 
\[
S_{\mathrm{HD}}\left( \varphi -\varphi \Delta _{n}Z_{\varphi },g,\lambda
-\Delta _{n}\lambda ,a-\Delta _{n}a,b-\Delta _{n}b,\kappa \right) , 
\]
which generates the additional counterterms 
\[
\Delta S_{m}\left( \varphi -\varphi \Delta _{n}Z_{\varphi },g,\lambda
-\Delta _{n}\lambda ,a-\Delta _{n}a,b-\Delta _{n}b,\kappa \right) . 
\]
Writing again $\Delta S_{m}=\Delta _{1}S_{m}+\Delta _{2}S_{m}$, we study the
effects of the redefinitions (\ref{redefa}) separately inside $\Delta
_{1}S_{m}$ and $\Delta _{2}S_{m}$. Inside $\Delta _{2}S_{m}$, the
redefinitions (\ref{redefa}) obviously generate terms of type $\Delta
_{2}S_{m}$, which can be removed renormalizing the couplings $\lambda
^{\prime }$ contained in $\Delta _{2}S_{m}$. Instead, inside $\Delta
_{1}S_{m}$ the redefinitions (\ref{redefa}) generate 
\begin{equation}
-\Delta _{n}\lambda \frac{\partial \Delta _{1}S_{m}}{\partial \lambda }%
-\varphi \Delta _{n}Z_{\varphi }\frac{\delta \Delta _{1}S_{m}}{\delta
\varphi }=-\int \mathrm{d}^{4}x\sqrt{-g}f_{\mu \nu }(g_{\rho \sigma })\left[
\Delta _{n}\lambda \frac{\partial T_{m}^{\mu \nu }}{\partial \lambda }%
+\varphi \Delta _{n}Z_{\varphi }\frac{\delta T_{m}^{\mu \nu }}{\delta
\varphi }\right] ,  \label{remnant}
\end{equation}
which are not of type $\Delta _{1}S_{m}$, in general. We keep the left-overs
(\ref{remnant}) in mind and proceed.

\bigskip

Now, consider the Feynman diagrams that do contain insertions of $\Delta
S_{m}$. The counterterms that contain at least two insertions of vertices of
type $\Delta _{1}S_{m}$ or one insertion of vertices of type $\Delta
_{2}S_{m}$ fall in the classes $\Delta _{2}S_{m}$ or $R_{\mu \nu }\mathcal{T}%
^{\mu \nu \rho \sigma }R_{\rho \sigma }$. Indeed, they are certainly
quadratic in the Ricci tensor, since they carry at least two Ricci tensors
on the external legs. Moreover, simple power counting shows that they are
made of matter operators that have at most dimensionality four, since the
matter operators contained in $S_{m}+\Delta S_{m}$ have at most
dimensionality four.

It remains to consider the Feynman diagrams that contain one vertex of type $%
\Delta _{1}S_{m}$ and no vertex of type $\Delta _{2}S_{m}$. Such diagrams
are collected in the expression 
\begin{equation}
\int \mathrm{d}^{4}x\sqrt{-g}f_{\mu \nu }(g_{\rho \sigma })\left\langle
T_{m}^{\mu \nu }\right\rangle _{0},  \label{asta}
\end{equation}
where 
\begin{equation}
\left\langle T_{m}^{\mu \nu }\right\rangle _{0}=-\frac{2}{\sqrt{-g}}\frac{%
\delta \Gamma _{0}[\Phi ,g_{\mu \nu }]}{\delta g_{\mu \nu }(x)}.
\label{asta2}
\end{equation}
Proceeding again by induction, assume that the divergences have been
renormalized up to the order $n$ included. Then the $(n+1)$-loop divergent
part of (\ref{asta}) is obtained replacing $\Gamma _{0}$ in (\ref{asta2})
with $\Gamma _{0\hspace{0.02in}\text{div}}^{(n)}$, given by (\ref{account}).
We thus see that the pure gravitational counterterms of (\ref{asta}) are
squarely proportional to the Ricci tensor or its covariant derivatives,
namely they belong to the purely gravitational sector of (\ref{saccadi}). On
the other hand, the matter contributions to the $(n+1)$-loop counterterms of
(\ref{asta}) are 
\[
-2\int \mathrm{d}^{4}xf_{\mu \nu }(g_{\rho \sigma })\left( \Delta
_{n}\lambda \frac{\partial }{\partial \lambda }+\varphi \Delta
_{n}Z_{\varphi }\frac{\delta }{\delta \varphi }\right) \frac{\delta S_{m}}{%
\delta g_{\mu \nu }}=\Delta _{n}\lambda \frac{\partial \Delta _{1}S_{m}}{%
\partial \lambda }+\varphi \Delta _{n}Z_{\varphi }\frac{\delta \Delta
_{1}S_{m}}{\delta \varphi }, 
\]

\noindent which are cancelled by the left-overs (\ref{remnant}). Observe
that the couplings $\lambda ^{\prime }$ contained in $\Delta _{1}S_{m}$ are
not renormalized. This is of course a consequence of the finiteness of the
energy-momentum tensor.

\bigskip

We have therefore proved that the theory (\ref{acca3}) of classical gravity
coupled with quantum matter is renormalizable in the form (\ref{acca3}). The
matter action $S_{m}+\Delta S_{m}$ is non-polynomial and $\Delta S_{m}$
contains infinitely many independent couplings. The set of independent
couplings, however, is considerably smaller than the set of independent
couplings of quantum gravity, since $\Delta S_{m}$ contains only lagrangian
terms of the form (\ref{first}).

\subsection{Non-vanishing cosmological constant}

When the theory contains masses and super-renormalizable parameters, the
cosmological constant is turned on by renormalization. The results just
proved generalize as follows. First, the non-renormalizable perturbation $%
\Delta _{1}S_{m}$ should be linearly proportional to the tensor 
\begin{equation}
\widehat{R}_{\mu \nu }=R_{\mu \nu }-\Lambda g_{\mu \nu }  \label{hat}
\end{equation}
or its covariant derivatives, instead of just the Ricci tensor. The hatted
tensor vanishes on the solutions of the field equations of the Einstein
action with a cosmological constant. Similarly, the vertices of type $\Delta
_{2}S_{m}$ should be squarely proportional to (\ref{hat}) and its covariant
derivatives. With these assumptions the perturbations will be written $%
\widehat{\Delta }_{1}S_{m}$ and $\widehat{\Delta }_{2}S_{m}$, respectively,
with $\widehat{\Delta }S_{m}=\widehat{\Delta }_{1}S_{m}+\widehat{\Delta }%
_{2}S_{m}$. Then the arguments of the previous subsection can be repeated to
prove the renormalizability of the action 
\[
S_{\mathrm{HD}}^{(\Lambda )}=\frac{1}{2\kappa ^{2}}\int \mathrm{d}^{4}x\sqrt{%
-g}\left( R-2\Lambda \right) +\int \mathrm{d}^{4}x\sqrt{-g}\widehat{R}_{\mu
\nu }\mathcal{T}^{\mu \nu \rho \sigma }\widehat{R}_{\rho \sigma
}+S_{m}(\varphi ,g,\lambda )+\widehat{\Delta }S_{m}(\varphi ,g,\lambda
,\lambda ^{\prime }). 
\]
The Newton constant $\kappa $ and the cosmological constant $\Lambda $ have
non-trivial renormalizations. In replacement of (\ref{aida}) the identity 
\begin{equation}
\frac{1}{2\kappa ^{2}}\int \mathrm{d}^{4}x\sqrt{-g}\left( R-2\Lambda \right)
+\int \mathrm{d}^{4}x\sqrt{-g}\widehat{R}_{\mu \nu }\mathcal{T}^{\mu \nu
\rho \sigma }\widehat{R}_{\rho \sigma }=\frac{1}{2\kappa ^{2}}\int \mathrm{d}%
^{4}x\sqrt{-g^{\prime }}\left( R(g^{\prime })-2\Lambda \right) .
\label{prove}
\end{equation}
can be applied to (\ref{prove}) to prove the renormalizability of the
acausal theory 
\[
S_{\text{AC}}^{(\Lambda )}=\frac{1}{2\kappa ^{2}}\int \mathrm{d}^{4}x\sqrt{-g%
}\left( R-2\Lambda \right) +S_{m}(\varphi ,g,\lambda )+\widehat{\Delta }%
S_{m}^{\prime }(\varphi ,g,\lambda ^{\prime }). 
\]

\section{More general renormalizable theories of acausal classical gravity
coupled with interacting quantum fields}

\setcounter{equation}{0}

The non-renormalizable perturbation $\Delta _{1}S_{m}$ (\ref{first}) of the
theories considered in the previous section has a special form. Precisely,
it contains a unique matter operator, the energy-momentum tensor $T_{m}^{\mu
\nu }$ of the unperturbed matter action $S_{m}$. In this section we prove
the renormalizability of more general theories.

In the higher-derivative models that we consider here, renormalization turns
on pure gravitational counterterms that do not fall in the class (\ref{spri}%
), so we need to study the map $\mathcal{M}$ for more general gravitational
actions. The most general modified action has the form 
\begin{equation}
S^{\prime }[\phi ]=S[\phi ]+S_{i}F_{ij}S_{j}+G_{i}S_{i}+H,  \label{sprin}
\end{equation}
where $G_{i}$ and $H$ are functions of the fields $\phi $, not necessarily
proportional to the field equations. It is not possible, in general, to find
a map that implements the equality (\ref{equa}). We describe what it is
convenient to do in this case.

Without loss of generality, we can assume that all kinetic terms are
contained in $S[\phi ]$ and $S_{i}F_{ij}S_{j}$, while $G_{i}S_{i}$ and $H$
are vertices. In the general case, for fields of spin smaller than or equal
to 2 in arbitrary dimension, this property follows for example from the
results of \cite{abse}. Here we refresh the proof for four-dimensional
gravity. For the purpose of studying the non-trivial counterterms induced by
renormalization, it is sufficient to work strictly in four dimensions.

Assume first that the cosmological constant vanishes and $S[\phi ]$ is the
Einstein action, so $S_{j}$ is proportional to the Ricci tensor. We wish to
prove that every scalar density that contains a non-trivial quadratic part
(when the metric is expanded in around flat space) is quadratic in the Ricci
tensor, up to vertex terms and total derivatives.

Every scalar density that contains a non-trivial quadratic part must
necessarily be of the form 
\begin{equation}
\sqrt{-g}R_{\mu \nu \rho \sigma }\nabla _{\lambda _{1}}\cdots \nabla
_{\lambda _{2n}}R_{\alpha \beta \gamma \delta },  \label{uno}
\end{equation}
with indices contracted in some way by means of the metric tensor. We
consider all possible contractions.

First consider the case in which some indices of one Riemann tensor are
contracted with themselves. Then (\ref{uno}) reduces to an expression of the
form 
\begin{equation}
\sqrt{-g}R_{\mu \nu \rho \sigma }\nabla _{\lambda _{1}}\cdots \nabla
_{\lambda _{2n}}R_{\alpha \beta }.  \label{ips}
\end{equation}
If no $\lambda $ is contracted with $\mu $, $\nu $, $\rho $ or $\sigma $,
then also the left Riemann tensor is actually a Ricci tensor or a curvature
scalar. If some $\lambda $ is contracted with an index of the Riemann
tensor, say $\mu $, then adding total derivatives and commuting the
covariant derivatives among themselves (which amounts to add vertex terms),
it is possible to move the derivative $\nabla ^{\mu }$ till it acts directly
on the Riemann tensor, obtaining 
\begin{equation}
\sqrt{-g}\nabla ^{\mu }R_{\mu \nu \rho \sigma }\nabla _{\lambda _{1}}\cdots
\nabla _{\lambda _{2n-1}}R_{\alpha \beta }.  \label{ipsilon}
\end{equation}
Then the Bianchi identity can be used to replace the Riemann tensor with a
second Ricci tensor.

Assume now that the indices of no Riemann tensor are contracted with
themselves. We can distinguish two cases: $i$) the case where at least one $%
\lambda $ is contracted with an index of a Riemann tensor, and $ii$) the
case where no $\lambda $ is contracted with an index of a Riemann tensor. In
case $i$) we can assume, eventually adding total derivatives, that the index 
$\lambda $ is contracted with an index of the right Riemann tensor, 
\begin{equation}
\sqrt{-g}R_{\mu \nu \rho \sigma }\nabla _{\lambda _{1}}\cdots \nabla
_{\lambda _{j-1}}\nabla ^{\lambda }\nabla _{\lambda _{j}}\cdots \nabla
_{\lambda _{2n-1}}R_{\lambda \beta \gamma \delta }.  \label{case}
\end{equation}
Then, adding vertex terms we can commute the covariant derivatives till $%
\nabla ^{\lambda }$ acts directly on the right Riemann tensor and use the
Bianchi identity to reduce (\ref{case}) to the form (\ref{ips}). In case $ii$%
) we can write, eventually after adding total derivatives and commuting
covariant derivatives, 
\begin{equation}
\sqrt{-g}R_{\mu \nu \rho \sigma }R^{\mu \nu \rho \sigma },\qquad \sqrt{-g}%
R_{\mu \nu \rho \sigma }\Box ^{2n-2}\nabla _{\alpha }\nabla ^{\alpha }R^{\mu
\nu \rho \sigma },  \label{iu}
\end{equation}
for $n=0$ and $n>0$, respectively. The $n=0$ case is dealt with using the
Gauss-Bonnet identity, as usual. The $n>0$ case is treated using the Bianchi
identity to write 
\[
-\sqrt{-g}R_{\mu \nu \rho \sigma }\Box ^{2n-2}\nabla _{\alpha }\left( \nabla
^{\sigma }R^{\mu \nu \alpha \rho }+\nabla ^{\rho }R^{\mu \nu \sigma \alpha
}\right) .
\]
Both terms have now the form (\ref{case}), so the arguments given above can
be applied again.

We have just considered gravity without a cosmological constant. Now we
generalize the result to the case in which $S[\phi ]$ is the Einstein action
with a cosmological constant. As before, it is convenient to use hatted
tensors 
\[
\widehat{R}_{\mu \nu \rho \sigma }=R_{\mu \nu \rho \sigma }-\frac{\Lambda }{3%
}(g_{\mu \rho }g_{\nu \sigma }-g_{\mu \sigma }g_{\nu \rho }) 
\]
and expand the metric around maximally symmetric solutions \cite{abse},
which have $\widehat{R}_{\mu \nu \rho \sigma }=0$. The scalar densities
contributing to the quadratic part in this expansion are classified studying
all possible contractions of 
\begin{equation}
\sqrt{-g}\widehat{R}_{\mu \nu \rho \sigma }\nabla _{\lambda _{1}}\cdots
\nabla _{\lambda _{2n}}\widehat{R}_{\alpha \beta \gamma \delta }.
\label{quat}
\end{equation}
Each such scalar density is quadratically proportional to the hatted Ricci
tensor $\widehat{R}_{\mu \nu }$, plus a linear combination of total
derivatives, vertices, $\sqrt{-g}R$ and $\sqrt{-g}$.

The proof follows the same arguments used above, replacing unhatted tensors
with hatted tensors everywhere. The only caveat is that the commutation of
two covariant derivatives generates the sum of a term containing an extra
hatted Riemann tensor, which is a vertex $\mathcal{O}\left( \widehat{R}%
^{3}\right) $, plus a term of type (\ref{quat}) with $n\rightarrow n-1$,
multiplied by an extra power of $\Lambda $. For example, on a vector $V_{\mu
}$%
\[
\lbrack \nabla _{\mu },\nabla _{\nu }]V_{\rho }=-R_{\mu \nu \rho \sigma
}V^{\sigma }=-\widehat{R}_{\mu \nu \rho \sigma }V^{\sigma }-\frac{\Lambda }{3%
}(g_{\mu \rho }V_{\nu }-g_{\nu \rho }V_{\mu }).
\]
The $\Lambda $-term lowers the number of curvature tensors, so the procedure
just described stops after a finite number of steps. The case $n=0$ is
treated using the identity 
\begin{equation}
\widehat{R}_{\mu \nu \rho \sigma }\widehat{R}^{\mu \nu \rho \sigma }=G_{B}+4%
\widehat{R}_{\mu \nu }\widehat{R}^{\mu \nu }-\widehat{R}^{2}-\frac{4}{3}%
\Lambda R+\frac{8}{3}\Lambda ^{2}.  \label{os}
\end{equation}

The integral of $\sqrt{-g}G_{B}$ is an irrelevant infinite additional
constant. The second and third terms on the right-hand side of (\ref{os})
are quadratically proportional to the field equations of $S[\phi ]$. The
forth and fifth terms on the right-hand side of (\ref{os}) give counterterms
that can be reabsorbed renormalizing the Newton constant and the
cosmological constant, respectively.

\bigskip

Coming back to (\ref{sprin}), we can thus assume that $G_{i}S_{i}$ and $H$
in (\ref{sprin}) are vertices and the kinetic terms are contained only in $%
S[\phi ]+S_{i}F_{ij}S_{j}$. We also assume that, in a weak field
approximation, $S_{i}=\mathcal{O}(\phi )$. If we apply the same map $\phi
^{\prime }(\phi )$ of section 2.2, which is such that $S[\phi
]+S_{i}F_{ij}S_{j}=S[\phi ^{\prime }]$, then $G_{i}S_{i}$ and $H$ are mapped
into vertices and we can write 
\begin{equation}
S^{\prime }[\phi ]=S[\phi ^{\prime }]+G_{i}^{\prime }[\phi ^{\prime
}]S_{i}[\phi ^{\prime }]+H^{\prime }[\phi ^{\prime }]  \label{esseprimo}
\end{equation}
for suitable new functions $G_{i}^{\prime }$ and $H^{\prime }$. The
undesirable higher-derivative kinetic terms have been removed. The
discussion about violations of causality is exactly the same as before,
since the map $\phi ^{\prime }(\phi )$ is the same.

Applying new maps $\mathcal{M}$ iteratively, we can eliminate every
newly-generated term that is quadratically proportional to $S_{i}$. To
isolate such new terms in the action (\ref{esseprimo}) rewrite it as 
\begin{equation}
S^{\prime \prime }[\phi ^{\prime }]\equiv S^{\prime }[\phi [\phi ^{\prime
}]]=S[\phi ^{\prime }]+S_{i}^{\prime }F_{ij}^{\prime }[\phi ^{\prime
}]S_{j}^{\prime }+\widetilde{G}_{i}^{\prime }[\phi ^{\prime }]S_{i}^{\prime
}+\widetilde{H}^{\prime }[\phi ^{\prime }],  \label{essesecondo}
\end{equation}
where $S_{i}^{\prime }=S_{i}[\phi ^{\prime }]$. Since the new terms are
certainly vertices, we have $\widetilde{F}_{ij}^{\prime }=\mathcal{O}(\phi
^{\prime })$. Applying a map $\mathcal{M}$ of the form $\phi ^{\prime \prime
}=\phi ^{\prime }+\mathcal{O}(\phi ^{\prime 2})$ we can reabsorb also $%
S_{i}^{\prime }F_{ij}^{\prime }S_{j}^{\prime }$ inside $S[\phi ^{\prime }]$.
In this way we end up with an $F_{ij}^{\prime \prime }$ that is $\mathcal{O}%
(\phi ^{\prime \prime 2})$. Repeating the procedure indefinitely, we reach $%
F_{ij}^{\infty }=0$, i.e. the complete elimination of the terms that are
quadratic in $S_{i}$. The final action reads 
\begin{equation}
S^{\infty }[\phi ^{\infty }]=S^{\prime }[\phi [\phi ^{\infty }]]=S[\phi
^{\infty }]+G_{i}^{\infty }[\phi ^{\infty }]S_{i}[\phi ^{\infty }]+H^{\infty
}[\phi ^{\infty }]  \label{esseinfinito}
\end{equation}
where $G_{i}^{\infty }$ and $H^{\infty }$ do not contain terms proportional
to $S_{i}[\phi ^{\infty }]$. Every $\phi ^{\infty }$-leg in $S^{\infty
}[\phi ^{\infty }]$ is multiplied by $1/\sqrt{1+2FS}$, because 
\[
\phi ^{\infty }=\sqrt{1+2FS}\phi +\mathcal{O}(\phi ^{2}),
\]
where $F$ and $S$ are the matrices $F_{ij}$ and $S_{ij}$ calculated at $\phi
=0$.

\bigskip

We can now use the results just derived to prove the renormalizability of a
more general class of acausal and higher-derivative theories. We assume
again, for simplicity, that the theories do not contain masses,
super-renormalizable couplings and the cosmological constant, because it is
straightforward to include them. Consider the higher-derivative theory 
\begin{equation}
S_{\mathrm{HD}}=\int \mathrm{d}^{4}x\sqrt{-g}\left[ \frac{R}{2\kappa ^{2}}+%
\overline{V}(g)+R_{\mu \nu }\mathcal{T}^{\mu \nu \rho \sigma }R_{\rho \sigma
}\right] +S_{m}(\varphi ,g,\lambda )+\int \mathrm{d}^{4}x\sqrt{-g}\sum_{I}%
\mathrm{O}_{I}(\varphi ,g)\overline{K}^{I}(g),  \label{esse}
\end{equation}
where $\overline{V}(g)$ is a vertex function of the metric, namely $%
\overline{V}(g)=\mathcal{O}(\phi ^{3})$ when the metric is expanded around
flat space, $g_{\mu \nu }=\eta _{\mu \nu }+2\kappa \phi _{\mu \nu }$.
Moreover, $S_{m}$ is the power-counting renormalizable matter action
embedded in curved space and $\mathrm{O}_{I}(\varphi ,g)$ is a basis of
covariant, gauge- (or BRST-) invariant local operators, not necessarily
scalar, of dimensionality smaller than or equal to four. The sources $%
\overline{K}^{I}(g)$ are arbitrary tensorial functions of the metric.

The renormalizability of (\ref{esse}) is easily proved. First, according to
the arguments of this section, the most general pure gravitational sector
without a cosmological constant can be written in the form appearing in (\ref
{esse}). In the notation (\ref{sprin}), $S$ is the Einstein action,while $%
\sqrt{-g}R_{\mu \nu }\mathcal{T}^{\mu \nu \rho \sigma }R_{\rho \sigma
}=S_{i}F_{ij}S_{j}$ and $\sqrt{-g}\overline{V}=G_{i}S_{i}+H$. Second, the
most general matter sector with operators of dimensionality smaller than or
equal to four, coupled with gravity in an arbitrary way, is precisely the
one appearing in (\ref{esse}). By power-counting, renormalization cannot
generate matter operators with dimensionalities greater than four, therefore
(\ref{esse}) is renormalizable.

Applying the map $\mathcal{M}$ to (\ref{esse}) we obtain the acausal theory 
\begin{equation}
S_{\mathrm{AC}}=\frac{1}{2\kappa ^{2}}\int \mathrm{d}^{4}x\sqrt{-g}\left[
R+V(g)\right] +S_{m}(\varphi ,g,\lambda )+\int \mathrm{d}^{4}x\sqrt{-g}%
\sum_{I}\mathrm{O}_{I}(\varphi ,g)K^{I}(g),  \label{acco}
\end{equation}
where $V(g)$ and $K^{I}(g)$ are other functions of the metric, and $V(g)$ is
a vertex. With the map $\mathcal{M}$ leading to (\ref{esseinfinito}), we can
assume that $V(g)$ does not contain terms quadratically proportional to the
Ricci tensor, or its covariant derivatives, but can contain terms that are
linearly proportional to it.

A simple example that illustrates the more general class of theories just
considered is the generalized acausal Einstein-Yang-Mills theory, which has
lagrangian 
\begin{equation}
\frac{\mathcal{L}_{\mathrm{EYM}\text{-}\mathrm{AC}}}{\sqrt{-g}}=\frac{1}{%
2\kappa ^{2}}R+V(g)-\frac{1}{4\alpha }\left\{ F_{\mu \nu }^{a}F^{a\hspace{%
0.01in}\mu \nu }~H(g)+T_{\mu \nu }K^{\mu \nu }(g)+\Upsilon _{\mu \nu \rho
\sigma }L^{\mu \nu \rho \sigma }(g)\right\} ,  \label{acca}
\end{equation}
where now $H(g)$, $K^{\mu \nu }(g)$ and $L^{\mu \nu \rho \sigma }(g)$ are
unconstrained functions of the metric and $V$ is a vertex function of the
metric.

\subsection{Consistent reductions of couplings}

The arbitrary functions contained in (\ref{acca}) can be restricted in
various ways, preserving renormalizability. A special class of theories are
those obtained from ordinary power-counting renormalizable theories in
curved space, once the couplings are allowed to depend on the spacetime
scalar curvature $R$. Then the subtraction of divergences can be performed
renormalizing the $R$-dependent couplings, and multiplying the fields with $%
R $-dependent wave-function renormalization constants, paying some attention
to the position of covariant derivatives. The lagrangian is uniquely
determined up to a few functions of $R$. The renormalization group, critical
exponents and beta functions are of course $R$-dependent.

We illustrate these facts in an explicit model. Consider the perturbed
higher-derivative Einstein-Yang-Mills theory 
\begin{eqnarray}
S_{\text{EYM-HD}}^{(f)} &=&\int \mathrm{d}^{n}x\sqrt{-g}\left[ \frac{1}{%
2\kappa ^{2}}R-\frac{1}{4\alpha }F_{\mu \nu }^{a}(L_{A}A)F^{a\hspace{0.01in}%
\mu \nu }(L_{A}A)\left( 1+\alpha Rf(R)\right) +\left( \xi +\gamma R\right)
W^{2}\right.  \nonumber \\
&&\left. +\left( \zeta +\rho R\right) G_{B}+\frac{\eta +\sigma R}{(n-1)^{2}}%
R^{2}+\frac{\tau }{n-1}R\Box R+R_{\mu \nu }\mathcal{T}_{4}^{\mu \nu \rho
\sigma }R_{\rho \sigma }\right] ,  \label{ostro}
\end{eqnarray}
where $f(R)$ and $L_{A}$ are power series in $R$, while $\gamma $, $\rho $, $%
\sigma $, $\tau $ are new coupling constants and $\mathcal{T}_{4}^{\mu \nu
\rho \sigma }$ is a (possibly differential) tensor operator such that $%
R_{\mu \nu }\mathcal{T}_{4}^{\mu \nu \rho \sigma }R_{\rho \sigma }$ is a
linear combination of the terms listed in appendix C, formula (\ref{eight}),
where $Q_{i}$, $i=1,\ldots 8,$ are arbitrary functions of the scalar
curvature $R$ with $Q_{j}(R)=\mathcal{O}(R^{2})$ for $j=1,2,3,4$. We prove
that (\ref{ostro}) is renormalizable and describe the acausal theory that is
obtained applying the map $\mathcal{M}$ to it. We also derive all order
formulas for the beta functions of $\gamma $, $\rho $, $\sigma $, $\tau $
and $f\equiv f(0)$, which are related to the trace-anomaly coefficients of
Yang-Mills theory in curved space. Note that (\ref{ostro}) contains a single
pure-gravitational term that is not quadratic in the Ricci tensor, that is
to say $RR^{\mu \nu \rho \sigma }R_{\mu \nu \rho \sigma }$, multiplied by
the coupling $\gamma +\rho $.

Clearly, the theory (\ref{ostro}) is nothing but Yang-Mills theory with an $%
R $-dependent gauge coupling. It contains a unique matter operator, $F^{2}$,
and is obtained perturbing the renormalizable higher-derivative
Einstein-Yang-Mills theory (\ref{basis}) with the vertex $RF^{2}$. The $R$%
-power series $f(R)$ and $L_{A}(R)$ and the pure gravitational vertices
listed in (\ref{eight}) are induced by renormalization.

\bigskip

To prove the renormalizability of (\ref{ostro}), it is convenient to start
from (\ref{basis}), deform it with $Rf(R)F^{2}$ and work perturbatively in $%
f(R)$. To the first order in $f(R)$ the Feynman diagrams coincide with those
of (\ref{basis}) with one insertion of the composite operator $F^{2}$. The
renormalization of $F^{2}$ in curved space\ has been studied by Hathrell in
QED \cite{hathrellQED}. Hathrell's derivation can be extended to non-Abelian
Yang-Mills theory in curved space using the Batalin-Vilkovisky formalism 
\cite{bata,wein}. First, the action (\ref{ostro}) is completed with a
gauge-fixing term and sources for the BRST\ transformations of the fields.
Then the BRST\ cohomology is extended to a $\sigma $-cohomology, defined by
the antiparenthesis with the complete action. Appendix C contains the
details and the complete proof of renormalizability, including the
characterization of the pure gravitational terms.

Hathrell's strategy (which we do not review here) can be applied in complete
analogy with the QED case. The result is formally identical to the one found
by Hathrell (formula (3.26) of \cite{hathrellQED}), apart from the $\sigma $%
-exact terms. Below we do not specify the $\sigma $-exact terms, which can
be dealt with using the formalism of appendix C. We obtain the renormalized
operator 
\begin{eqnarray}
\frac{1}{4\alpha }[F_{\mu \nu }^{a}F^{a\hspace{0.01in}\mu \nu }] &=&-\frac{%
\varepsilon \alpha \mu ^{\varepsilon }}{4\widehat{\beta }\alpha _{\text{B}}}%
F_{\text{B}\mu \nu }^{a}F_{\text{B}}^{a\hspace{0.01in}\mu \nu }+\frac{\alpha 
}{\widehat{\beta }}\left[ \left( \varepsilon L_{\xi }-\beta _{\xi }\right)
W^{2}+\left( \varepsilon L_{\zeta }-\beta _{\zeta }\right) G_{B}\right.  
\nonumber \\
&&\left. +\left( \varepsilon L_{\eta }-\beta _{\eta }\right) \frac{R^{2}}{%
(n-1)^{2}}+\frac{4}{n-1}\left( L_{\eta }-\frac{\alpha \beta _{\eta }}{\beta }%
\right) \Box R\right] +\sigma X,  \label{FF}
\end{eqnarray}
where the subscript B denotes the bare quantities. We can immediately read
the renormalization constant $Z_{F^{2}}$ of the operator $F^{2}$ in flat
space, 
\[
Z_{F^{2}}=Z_{\alpha }\left( 1-\frac{\beta (\alpha )}{\varepsilon \alpha }%
\right) 
\]
and the $f$-beta function 
\begin{equation}
f_{\mathrm{B}}=\mu ^{-\varepsilon }fZ_{F^{2}}^{-1},\qquad \frac{\mathrm{d}f}{%
\mathrm{d}\ln \mu }=\varepsilon f+\beta _{f},\qquad \beta _{f}=f\frac{%
\mathrm{d}\ln Z_{F^{2}}}{\mathrm{d}\ln \mu }=f\left( \beta ^{\prime }-\frac{%
2\beta }{\alpha }\right) ,  \label{f2}
\end{equation}
where the prime denotes differentiation with respect to the gauge coupling $%
\alpha $.

Comparing the terms of dimensionality six in the bare and renormalized
actions, we can derive the beta functions of $\gamma $, $\rho $, $\sigma $
and $\tau $. Factorizing one power of $R$, we have 
\begin{eqnarray*}
- &&f_{\mathrm{B}}\frac{1}{4}F_{\mathrm{B}\mu \nu }^{a}F_{\mathrm{B}}^{a%
\hspace{0.01in}\mu \nu }+\gamma _{\mathrm{B}}W^{2}+\rho _{\mathrm{B}}G_{B}+%
\frac{\sigma _{\mathrm{B}}}{(n-1)^{2}}R^{2}+\frac{\tau _{\mathrm{B}}}{n-1}%
\Box R= \\
&&-f\frac{\mu ^{-\varepsilon }}{4}[F_{\mu \nu }^{a}F^{a\hspace{0.01in}\mu
\nu }]+\gamma \mu ^{-\varepsilon }W^{2}+\rho \mu ^{-\varepsilon }G_{B}+\frac{%
\sigma \mu ^{-\varepsilon }}{(n-1)^{2}}R^{2}+\frac{\tau \mu ^{-\varepsilon }%
}{n-1}\Box R,
\end{eqnarray*}
up to $\sigma $-exact terms. Using (\ref{FF}) we find 
\begin{eqnarray*}
\gamma _{\mathrm{B}} &=&\gamma \mu ^{-\varepsilon }+\mu ^{-\varepsilon }f%
\frac{\alpha ^{2}}{\widehat{\beta }}\left( \beta _{\xi }-\varepsilon L_{\xi
}\right) ,\qquad \rho _{\mathrm{B}}=\rho \mu ^{-\varepsilon }+\mu
^{-\varepsilon }f\frac{\alpha ^{2}}{\widehat{\beta }}\left( \beta _{\zeta
}-\varepsilon L_{\zeta }\right) , \\
\sigma _{\mathrm{B}} &=&\sigma \mu ^{-\varepsilon }+\mu ^{-\varepsilon }f%
\frac{\alpha ^{2}}{\widehat{\beta }}\left( \beta _{\eta }-\varepsilon
L_{\eta }\right) ,\qquad \tau _{\mathrm{B}}=\tau \mu ^{-\varepsilon }+\mu
^{-\varepsilon }4f\frac{\alpha ^{2}}{\widehat{\beta }}\left( \frac{\alpha
\beta _{\eta }}{\beta }-L_{\eta }\right) .
\end{eqnarray*}
Differentiating with respect to $\ln \mu $ and simplifying by means of (\ref
{abeta}), we immediately derive the beta functions 
\begin{eqnarray*}
\widehat{\beta }_{\gamma } &=&\frac{\mathrm{d}\gamma }{\mathrm{d}\ln \mu }%
=\varepsilon \gamma -f\alpha ^{2}\frac{\mathrm{d}\beta _{\xi }}{\mathrm{d}%
\alpha },\qquad \widehat{\beta }_{\rho }=\varepsilon \rho -f\alpha ^{2}\frac{%
\mathrm{d}\beta _{\zeta }}{\mathrm{d}\alpha }, \\
\widehat{\beta }_{\sigma } &=&\varepsilon \sigma -f\alpha ^{2}\frac{\mathrm{d%
}\beta _{\eta }}{\mathrm{d}\alpha },\qquad \widehat{\beta }_{\tau
}=\varepsilon \tau -4f\alpha \frac{\mathrm{d}}{\mathrm{d}\alpha }\left( 
\frac{\alpha ^{2}\beta _{\eta }}{\beta }\right) .
\end{eqnarray*}

Next, consider the Feynman diagrams of higher order in $f$, obtained with
multiple insertions of $RF^{2}$ and its BRST completion $R\sigma X$, where $%
X $ is the unspecified functional appearing in (\ref{FF}). Diagrams with no
gauge fields, ghosts and BRST\ sources on the external legs contribute to
the pure gravitational counterterms. Since every insertion of $RF^{2}$ plus
its completion carries a factor of $R$, the pure gravitational counterterms
with multiple insertions are certainly squarely proportional to the Ricci
tensor and can be renormalized redefining the couplings contained in $%
\mathcal{T}_{4}^{\mu \nu \rho \sigma }$ (whose form is so far unspecified).
BRST invariance, parity invariance and power counting ensure that the
divergent diagrams that carry gauge fields, ghost and BRST\ sources on the
external legs can only give counterterms proportional to $F^{2}$, plus $%
\sigma $-exact contributions, and carry a power of $R$ at least equal to the
number of insertions. The $\sigma $-exact contributions contain also
gauge-field redefinitions, that can only have the form $A_{\mu
}^{a}\rightarrow P(R)A_{\mu }^{a}$, for a certain function $P$. These
divergences can be subtracted renormalizing the parameters that are
contained in the functions $f(R)$ and $L_{A}$ of (\ref{ostro}) and in the
BRST-exact sector added to (\ref{ostro}). The pure gravitational
contributions are further restricted by the analysis of appendix C to a
linear combination of eight independent terms, multiplied by arbitrary
functions of the scalar curvature.

\bigskip

Finally, we can read the renormalizable acausal Einstein-Yang-Mills theory
associated with (\ref{ostro}) applying the map $\mathcal{M}$ in the form (%
\ref{esseinfinito}). Such a map reabsorbs every term that is quadratically
proportional to the Ricci tensor. There exists a function $\overline{g}(g)$
such that if we replace $g$ with $\overline{g}(g)$ in (\ref{ostro}) we
obtain an acausal action of the form 
\begin{eqnarray}
S_{\text{EYM-AC}}(g) &=&\int \mathrm{d}^{4}x\sqrt{-g}\left[ \frac{R}{2\kappa
^{2}}+(\gamma +\rho )\overline{RR_{\mu \nu \rho \sigma }R^{\mu \nu \rho
\sigma }}\right]   \label{sacca} \\
&&-\frac{1}{4\alpha }\int \mathrm{d}^{4}x\ \sqrt{-\overline{g}}F_{\mu \nu
}^{a}(L_{A}(\overline{g})A)F_{\rho \sigma }^{a}(L_{A}(\overline{g})A)%
\overline{g}^{\mu \rho }\overline{g}^{\nu \sigma }\left[ \ 1+\alpha R(%
\overline{g})f(R(\overline{g}))\right] .  \nonumber
\end{eqnarray}
Here $\overline{g}$ should be understood as the given function $\overline{g}%
(g)$ and $\sqrt{-g}\overline{RR_{\mu \nu \rho \sigma }R^{\mu \nu \rho \sigma
}}$ denotes the terms generated by the vertex $\sqrt{-\overline{g}}R(%
\overline{g})R_{\mu \nu \rho \sigma }(\overline{g})R^{\mu \nu \rho \sigma }(%
\overline{g})$, once every contribution quadratically proportional to the
Ricci tensor is eliminated by the iterative procedure explained in the
previous subsection. In the end, the surviving vertex has basically the same
form as $RR_{\mu \nu \rho \sigma }R^{\mu \nu \rho \sigma }$, but every $\phi 
$-leg gets multiplied by the function $1/\sqrt{1+2FS}$ of (\ref{cebiso}) and
(\ref{cnf}), which is responsible for the classical violations of causality.

The theory (\ref{sacca}) is more general than the ones considered in section
4, because the list $\Delta S^{\text{(HEAD)}}$ of vertices that have
dimensionality six does not contain just the stress tensor multiplied by the
Ricci tensor, but also the vertex $RF^{2}$ and the pure gravitational vertex 
$RR^{\mu \nu \rho \sigma }R_{\mu \nu \rho \sigma }$. Precisely, 
\[
\Delta S^{\text{(HEAD)}}=\int \mathrm{d}^{4}x\sqrt{-g}\left[ -\frac{f}{4}%
F_{\mu \nu }^{a}F^{a\hspace{0.01in}\mu \nu }R-\frac{a}{2}R_{\mu \nu }T^{\mu
\nu }+(\gamma +\rho )RR^{\mu \nu \rho \sigma }R_{\mu \nu \rho \sigma
}\right] . 
\]
On the other hand, the theory is less general than (\ref{acca}), because the
arbitrariness contained in (\ref{acca}) is reduced in (\ref{sacca}) to a few
functions of $R$.

The arguments of this section can be easily applied to classical Einstein
gravity coupled with quantum electrodynamics with an $R$-dependent electric
charge, or any other power-counting renormalizable quantum field theory with 
$R$-dependent couplings.

\section{Conclusions}

\setcounter{equation}{0}

In this paper we have studied new renormalizable acausal theories of
classical gravity coupled with interacting quantum fields. Performing
systematic field redefinitions of the metric tensor, the divergences are
removed without introducing higher-derivative kinetic terms in the
gravitational sector. Previous results are generalized and new theorems are
proved. In particular, it is shown how to treat quantized fields which
interact through $R$-dependent vertices.

We have studied Einstein-Yang-Mills theory with an $R$-dependent gauge
coupling in detail. The perturbation $RF^{2}$ induces extra gravitational
terms, one of which, $RR_{\mu \nu \rho \sigma }R^{\mu \nu \rho \sigma }$, is
not squarely proportional to the Ricci tensor. General formulas for the beta
functions of the vertices with dimensionality six are derived. They are
expressed in terms of the trace-anomaly coefficients of the matter sector
embedded in curved background. The renormalization-group flow depends on the
scalar curvature of spacetime.

Our results can be extended to all power-counting renormalizable quantum
field theories with $R$-dependent coupling constants coupled with classical
gravity. More general renormalizable theories contain all matter operators
with dimensionalities smaller than or equal to four, multiplied by arbitrary
tensorial functions of the metric. In complete generality, the pure
gravitational counterterms that contain higher-derivative kinetic
contributions can be removed using the map $\mathcal{M}$, which trades the
instabilities due to higher-derivatives for classical violations of
causality.

The causality violations introduced by the map $\mathcal{M}$ are governed by
the parameters $a$ and $b^{\prime }$, and can be detected at high energies.
The renormalization-group flow of $a$ and $b^{\prime }$ can be studied
exactly in a large class of models, for example those that interpolate
between UV\ and IR\ conformal fixed points.

\vskip 20truept \noindent {\Large \textbf{Appendices}}

\vskip 20truept \noindent {\Large \textbf{A\quad Working without bitensors}}%
\vskip 10truept

\renewcommand{\theequation}{A.\arabic{equation}} \setcounter{equation}{0}

In the case of gravity the functional derivatives $S_{ijk\cdots }$ are bi-,
tri-tensor densities, and so on, and they involve several delta functions.
It is unpleasant to work with these objects, but it is not strictly
necessary for our purposes. Here we show how to use the map $\mathcal{M}$
working only with tensors and tensor densities. Write 
\begin{equation}
S^{\prime }[g_{i}]=S[g_{i}]+\int_{ij}\sqrt{-g}S_{i}F_{ij}S_{j},
\label{sprimo}
\end{equation}
where $F_{ij}$ is symmetric and can contain derivative operators. The index $%
i$ labels also the spacetime point, as usual. Now the summation-integration
over repeated indices is not understood, but written explicitly on the
integral sign as shown in (\ref{sprimo}). Define the field-equation tensor 
\[
S_{i}=\frac{\widetilde{\delta }S}{\widetilde{\delta }g_{i}},\qquad \frac{%
\widetilde{\delta }}{\widetilde{\delta }g_{i}}\equiv \frac{1}{\sqrt{-\det
g_{i}}}\frac{\delta }{\delta g_{i}}. 
\]
Then, there exists a field redefinition 
\begin{equation}
g_{i}^{\prime }=g_{i}+\int_{j}\Delta _{ij}S_{j},  \label{gprimo}
\end{equation}
such that, perturbatively in $F$ and to all orders in powers of $F$, 
\begin{equation}
S^{\prime }[g_{i}]=S[g_{i}^{\prime }].  \label{theorem}
\end{equation}
The condition (\ref{theorem}) can be written as 
\[
S[g_{i}]+\int_{ij}\sqrt{-g}S_{i}F_{ij}S_{j}=S\left[ g_{i}+\int_{j}\Delta
_{ij}S_{j}\right] =S[g_{i}]+\sum_{n=1}^{\infty }\frac{1}{n!}\int\Sb %
k_{1}\cdots k_{n}  \\ m_{1}\cdots m_{n}  \endSb \widetilde{S}_{k_{1}\cdots
k_{n}}\prod_{l=1}^{n}(\Delta _{k_{l}m_{l}}S_{m_{l}}), 
\]
after a Taylor expansion, where $\widetilde{S}_{k_{1}\cdots k_{n}}\equiv
\delta ^{n}S/(\delta g_{k_{1}}\cdots \delta g_{k_{n}})$ if an $n$-tensor
density. The equality is verified if 
\begin{equation}
\Delta _{ij}=F_{ij}-\frac{1}{\sqrt{-\det g_{i}}}\int\Sb k_{1}\cdots k_{n} 
\\ m_{3}\cdots m_{n}  \endSb \Delta _{k_{1}i}\Delta
_{k_{2}j}\sum_{n=2}^{\infty }\frac{1}{n!}\widetilde{S}_{k_{1}k_{2}k_{3}%
\cdots k_{n}}\prod_{l=3}^{n}(\Delta _{k_{l}m_{l}}S_{m_{l}}),  \label{inequa}
\end{equation}
where the product is meant to be 1 for $n=2$.

Let $A_{i}$ be tensor and 
\[
F=\frac{1}{2}\int_{ij}\sqrt{-g_{i}}A_{i}F_{ij}A_{j} 
\]
the scalar quadratic form defined by $F_{ij}$. Then 
\begin{equation}
B_{i}\equiv \int_{j}F_{ij}A_{j}=\frac{\widetilde{\delta }F}{\widetilde{%
\delta }A_{i}}  \label{auto}
\end{equation}
is a tensor. Moreover, the symmetry property of $F_{ji}$ reads 
\begin{equation}
\frac{1}{\sqrt{-\det g_{i}}}F_{ji}=F_{ij}\frac{1}{\sqrt{-\det g_{j}}}.
\label{symm}
\end{equation}
Observe that (\ref{inequa}) implies that $\Delta _{ij}$ satisfies the same
symmetry property:

\begin{equation}
\frac{1}{\sqrt{-\det g_{i}}}\Delta _{ji}=\Delta _{ij}\frac{1}{\sqrt{-\det
g_{j}}}.  \label{symm2}
\end{equation}

Although $F_{ij}$, $\Delta _{ij}$ and $\widetilde{S}_{ij}$ are bitensors, we
do not really need to use them explicitly. For example, 
\begin{equation}
C_{k_{1}}\equiv \frac{1}{\sqrt{-\det g_{k_{1}}}}\int_{k_{2}}\widetilde{S}%
_{k_{1}k_{2}}B_{k_{2}}  \label{ck1}
\end{equation}
is a tensor that can be calculated as 
\begin{equation}
C_{k_{1}}=\frac{1}{\sqrt{-\det g_{k_{1}}}}\int_{k_{2}}\frac{\delta 
\widetilde{S}_{k_{1}}}{\delta g_{k_{2}}}B_{k_{2}}=\frac{1}{\sqrt{-\det
g_{k_{1}}}}\left. \delta \widetilde{S}_{k_{1}}(\delta g)\right| _{\delta
g=B},  \label{operation1}
\end{equation}
without working out the bi-tensor density $S_{k_{1}k_{2}}$ explicitly. Since 
$\widetilde{S}_{k_{1}}$ is a tensor density, also $\left. \delta \widetilde{S%
}_{k_{1}}(\delta g)\right| _{\delta \phi =B}$ is a tensor density, so $%
C_{k_{1}}$ is a tensor.

Define the tensors 
\[
E_{i}^{(n)}=\int_{j}F_{ij}D_{j}^{(n)}\qquad D_{i}^{(n)}=\frac{1}{\sqrt{-\det
g_{i}}}\int_{j}\widetilde{S}_{ij}E_{j}^{(n-1)}, 
\]
recursively calculable from $D_{i}^{(0)}\equiv S_{i}$, with a chain of
operations (\ref{operation1}) and (\ref{auto}), and the tensor dentities 
\[
\widetilde{D}_{i}^{(n)}=\sqrt{-\det g_{i}}D_{i}^{(n)}. 
\]
So 
\begin{eqnarray}
g_{i}^{\prime } &=&g_{i}+\Delta _{ij}S_{j}=g_{i}+E_{i}^{(0)}-\frac{1}{2}%
E_{i}^{(1)}+\frac{1}{2}E_{i}^{(2)}  \label{trasfa} \\
&&-\frac{1}{3!}\frac{1}{\sqrt{-\det g_{i}}}\int_{k_{1},k_{2},k_{3}}F_{k_{1}i}%
\widetilde{S}_{k_{1}k_{2}k_{3}}E_{k_{2}}^{(0)}E_{k_{3}}^{(0)}+\mathcal{O}%
(F^{4}).  \nonumber
\end{eqnarray}
The first line is immediate to calculate, with the operations already given.
The second line is less immediate. Write 
\begin{equation}
\int_{k_{2},k_{3}}\widetilde{S}%
_{k_{1}k_{2}k_{3}}E_{k_{2}}^{(n)}E_{k_{3}}^{(m)}=\int_{k_{3}}\frac{\delta 
\widetilde{D}_{k_{1}}^{(n+1)}}{\delta g_{k_{3}}}E_{k_{3}}^{(m)}-%
\int_{k_{2},k_{3}}\widetilde{S}_{k_{1}k_{2}}\frac{\delta E_{k_{2}}^{(n)}}{%
\delta g_{k_{3}}}E_{k_{3}}^{(m)}  \label{due}
\end{equation}
In this form, the second line of (\ref{trasfa}) can be calculated
straightforwardly, using the same procedure as for (\ref{operation1}).
Clearly (\ref{due}) a tensor density and the second line of (\ref{trasfa})
is a tensor.

At the fourth order, we find objects such as 
\[
\int_{k_{2},k_{3}}\widetilde{S}%
_{k_{1}k_{2}k_{3}}E_{k_{2}}^{(1)}E_{k_{3}}^{(0)},\qquad
\int_{k_{2},k_{3},k_{4}}\widetilde{S}%
_{k_{1}k_{2}k_{3}k_{4}}E_{k_{2}}^{(0)}E_{k_{3}}^{(0)}E_{k_{4}}^{(0)}, 
\]
etc. The first of these can be commuted using (\ref{due}). The second can be
computed writing 
\[
\int_{k_{2},k_{3},k_{4}}\widetilde{S}%
_{k_{1}k_{2}k_{3}k_{4}}E_{k_{2}}^{(0)}E_{k_{3}}^{(0)}E_{k_{4}}^{(0)}=%
\int_{k_{2},k_{3},k_{4}}\frac{\delta (\widetilde{S}%
_{k_{1}k_{2}k_{3}}E_{k_{2}}^{(0)}E_{k_{3}}^{(0)})}{\delta g_{k_{4}}}%
E_{k_{4}}^{(0)}-2\int_{k_{2},k_{3},k_{4}}\widetilde{S}%
_{k_{1}k_{2}k_{3}}E_{k_{2}}^{(0)}\frac{\delta E_{k_{3}}^{(0)}}{\delta
g_{k_{4}}}E_{k_{4}}^{(0)}, 
\]
and so on. The procedure can be iterated to all orders.

\vskip 20truept {\Large \noindent \textbf{B\quad The map M for gravity}}%
\vskip 10truept

\renewcommand{\theequation}{B.\arabic{equation}} \setcounter{equation}{0}

We work out the map $g^{\prime }(g,a,b)$ such that 
\[
S^{\prime }[g]\equiv \int \mathrm{d}^{n}x\sqrt{-g}\left[ R+aR_{\mu \nu
}R^{\mu \nu }+bR^{2}\right] =\int \mathrm{d}^{n}x\sqrt{-g^{\prime }}%
R(g^{\prime })\equiv S[g^{\prime }] 
\]
to the orders $a^{2}$, $ab$, $b^{2}$, in generic spacetime dimension $n$. We
have 
\[
S=\frac{1}{2\kappa ^{2}}\int \mathrm{d}^{n}x\sqrt{-g}R(g),\qquad
D_{i}^{(0)}=S_{i}=-\frac{1}{2\kappa ^{2}}\left( R^{\mu _{i}\nu _{i}}-\frac{1%
}{2}g^{\mu _{i}\nu _{i}}R\right) , 
\]
where $i=(\mu _{i},\nu _{i},x_{i})$ collects the spacetime indices and the
spacetime point and $g_{i}=g_{\mu _{i}\nu _{i}}(x_{i})$. Next, we have 
\[
F_{ij}=2\kappa ^{2}\left\{ \frac{a}{2}(g_{\mu _{i}\mu _{j}}g_{\nu _{i}\nu
_{j}}+g_{\mu _{i}\nu _{j}}g_{\nu _{i}\mu _{j}})+\overline{b}g_{\mu _{i}\nu
_{i}}g_{\mu _{j}\nu _{j}}\right\} \delta ^{(n)}(x_{i}-x_{j}),\qquad 
\overline{b}=\frac{4b+a(4-n)}{(n-2)^{2}}. 
\]
Multiplying $F_{ij}$ by $S_{i}$ we find. 
\[
E_{i}^{(0)}(g)=-aR_{\mu _{i}\nu _{i}}+\frac{v}{2}g_{\mu _{i}\nu
_{i}}R,\qquad v=a+(n-2)\overline{b}. 
\]
Next, applying (\ref{due}) we get 
\begin{eqnarray*}
E_{i}^{(1)}(g) &=&\frac{a^{2}}{2}\Box R_{\mu _{i}\nu _{i}}-\frac{av}{4}%
(n-2)\nabla _{\mu _{i}}\nabla _{\nu _{i}}R-\frac{a}{4}\left(
(n-4)v-4a\right) RR_{\mu _{i}\nu _{i}}-a^{2}R_{\mu _{i}}^{\lambda
}R_{\lambda \nu _{i}} \\
&&-a^{2}R_{\mu _{i}\alpha \nu _{i}\beta }R^{\alpha \beta }+g_{\mu _{i}\nu
_{i}}\left\{ \frac{v}{4}\left( (n-1)v-2a\right) \Box R+\frac{a(v-2\overline{b%
})}{2}R_{\alpha \beta }R^{\alpha \beta }+uR^{2}\right\} ,\qquad \qquad
\end{eqnarray*}
where 
\[
u=\frac{(n-4)}{8}\left( v^{2}-2a\overline{b}\right) -\frac{a^{2}}{4}. 
\]
The map $\mathcal{M}$ reads, to the second order in $F$, 
\[
g_{i}^{\prime }=g_{i}+\Delta _{ij}S_{j}=g_{i}+E_{i}^{(0)}-\frac{1}{2}%
E_{i}^{(1)}+\mathcal{O}(F^{3}). 
\]
The inverse map reads 
\[
g_{i}=g_{i}^{\prime }-E_{i}^{(0)}(g^{\prime })+\left. \delta
E_{i}^{(0)}(g^{\prime })\right| _{\delta g^{\prime }=E^{(0)}(g^{\prime })}+%
\frac{1}{2}E_{i}^{(1)}(g^{\prime })+\mathcal{O}(F^{3}), 
\]
where, to the same order, 
\begin{eqnarray*}
\left. \delta E_{i}^{(0)}(g^{\prime })\right| _{\delta g^{\prime
}=E^{(0)}(g^{\prime })} &=&E_{i}^{(1)}(g^{\prime })+2a^{2}R_{\mu
_{i}}^{\lambda }R_{\lambda \nu _{i}}+\frac{a}{4}\left( (n-6)v-4a\right)
RR_{\mu _{i}\nu _{i}} \\
&&+g_{\mu _{i}\nu _{i}}\left( a\overline{b}R_{\alpha \beta }R^{\alpha \beta
}-uR^{2}\right) .
\end{eqnarray*}

\vskip 20truept \noindent {\Large \textbf{C\quad Batalin-Vilkovisky
formalism for perturbed Einstein-Yang-Mills theory}}\vskip 10truept

\renewcommand{\theequation}{C.\arabic{equation}} \setcounter{equation}{0}

When gauge couplings depend on spacetime, the fields and their BRST
transformations are renormalized in a non-trivial way. To keep track of such
renormalizations it is necessary to use the Batalin-Vilkovisky formalism 
\cite{bata,wein}. We focus on higher-derivative Einstein-Yang-Mills theory
and perturb it with the vertex $fRF^{2}$.

Consider first the unperturbed case $f=0$. We introduce Faddeev-Popov ghosts 
$C^{a}$, antighosts $\overline{C}^{a}$ and a Lagrange multiplier $B^{a}$ for
the gauge-fixing. The fields are collectively denoted with $\Phi
^{i}=(A_{\mu }^{a},\overline{C}^{a},C^{a},B^{a})$. We add BRST sources $%
K_{i}=(K_{a}^{\mu },K_{\overline{C}}^{a},K_{C}^{a},K_{B}^{a})$ for every
field $\Phi ^{i}$ and extend the action (\ref{basis}) as 
\begin{eqnarray}
\mathcal{S}_{\text{EYM-HD}} &=&\int \mathrm{d}^{n}x\sqrt{-g}\left[ \frac{R}{%
2\kappa ^{2}}+\xi W^{2}+\zeta G_{B}+\frac{\eta }{(n-1)^{2}}R^{2}-\frac{1}{%
4\alpha }F_{\mu \nu }^{a}F^{a\hspace{0.01in}\mu \nu }\right]  \label{act} \\
&&+s\Psi (\Phi ,g)-\int \mathrm{d}^{n}x\sqrt{-g}\left[ \left( sA_{\mu
}^{a}\right) K_{a}^{\mu }+\left( s\overline{C}^{a}\right) K_{\overline{C}%
}^{a}+\left( sC^{a}\right) K_{C}^{a}+\left( sB^{a}\right) K_{B}^{a}\right] ,
\nonumber
\end{eqnarray}
where the BRST\ transformations are 
\[
sA_{\mu }^{a}=\partial _{\mu }C^{a}+f^{abc}A_{\mu }^{b}C^{c},\qquad sC^{a}=-%
\frac{1}{2}f^{abc}C^{b}C^{c},\qquad s\overline{C}^{a}=B^{a},\qquad sB^{a}=0. 
\]
We choose the gauge fixing $\nabla ^{\mu }A_{\mu }^{a}=0$, which breaks the
gauge symmetry but preserves general covariance: 
\begin{equation}
\Psi (\Phi ,g)\equiv \int \mathrm{d}^{n}x\sqrt{-g}\left[ -\frac{\lambda }{2}%
\overline{C}^{a}B^{a}+\overline{C}^{a}\nabla ^{\mu }A_{\mu }^{a}\right] .
\label{psi}
\end{equation}
Define the antiparenthesis 
\begin{equation}
(X,Y)=\int \mathrm{d}^{n}x\sqrt{-g(x)}\left\{ \frac{\widetilde{\delta }_{r}X%
}{\widetilde{\delta }\Phi ^{i}(x)}\frac{\widetilde{\delta }_{l}Y}{\widetilde{%
\delta }K_{i}(x)}-\frac{\widetilde{\delta }_{r}X}{\widetilde{\delta }K_{i}(x)%
}\frac{\widetilde{\delta }_{l}Y}{\widetilde{\delta }\Phi ^{i}(x)}\right\} ,
\label{antipar}
\end{equation}
where the tilded derivatives are normal derivatives divided by $\sqrt{-g}$.
A canonical transformation of fields and sources is a transformation that
preserves the antiparenthesis. It is generated by a functional $\mathcal{F}%
(\Phi ,K^{\prime })$ and reads 
\[
\Phi ^{i\ \prime }=\frac{\widetilde{\delta }\mathcal{F}}{\widetilde{\delta }%
K_{i}^{\prime }},\qquad K_{i}=\frac{\widetilde{\delta }\mathcal{F}}{%
\widetilde{\delta }\Phi ^{i}}, 
\]
The generating functional of the identity transformation is 
\[
I(\Phi ,K^{\prime })=\int \mathrm{d}^{n}x\sqrt{-g}\sum_{i}\Phi
^{i}K_{i}^{\prime }. 
\]
The BRST\ invariance is generalized to the identity 
\begin{equation}
(\mathcal{S}_{\text{EYM-HD}},\mathcal{S}_{\text{EYM-HD}})=0.  \label{nil}
\end{equation}
Define also the generalized BRST operator 
\begin{equation}
\sigma X\equiv (\mathcal{S}_{\text{EYM-HD}},X),  \label{sigma}
\end{equation}
which is nilpotent, $\sigma ^{2}=0$, because of the identity (\ref{nil}).

The identity (\ref{nil}) ensures also, in every regularization scheme, 
\begin{equation}
(\Gamma _{\text{EYM-HD}},\Gamma _{\text{EYM-HD}})=0,  \label{milpo}
\end{equation}
where $\Gamma _{\text{EYM-HD}}$ is the generating functional of one-particle
irreducible diagrams, including the diagrams that have ghosts, Langrange
multipliers and BRST-sources on their external legs.

Proceeding inductively, assume that $\Gamma _{\text{EYM-HD}}^{\ (n)}$ is the
generating functional renormalized up to the $n$-th loop included. General
renormalization theory and (\ref{milpo}) ensure that the ($n+1$)-loop
divergences $\Gamma _{\text{div}}^{\ (n+1)}$ are local and $\sigma $-closed, 
$\sigma \Gamma _{\text{div}}^{\ (n+1)}=0$. The most general solution to this
condition has the form 
\begin{equation}
\Gamma _{\text{div}}^{\ (n+1)}=\mathcal{G}_{n}+\sigma \mathcal{R}_{n},
\label{counter}
\end{equation}
where $\mathcal{G}_{n}$ is a gauge-invariant functional depending only on $%
A_{\mu }^{a}$ (and the metric tensor). By parity invariance, $\mathcal{G}_{n}
$ can only be the sum of a term proportional to $F^{2}$, which is reabsorbed
renormalizing the squared gauge coupling $\alpha $, plus pure gravitational
terms with dimensionality four, that are reabsorbed renormalizing $\xi $, $%
\zeta $, $\eta $ and $\tau $. The $\sigma $-exact terms can be easily
classified. They are equal to $\sigma $ acting on a functional with
dimensionality three and ghost number $-1$. For future use, we list all of
the scalars, vectors and tensors that have dimensionality three or less and
ghost number $-1$. There are 17 such objects: 
\begin{eqnarray}
&&K_{a}^{\mu }A_{\nu }^{a},\quad K_{C}^{a}C^{a},\quad K_{\overline{C}}^{a}%
\overline{C}^{a},\quad K_{B}^{a}B^{a},\quad \overline{C}^{a}B^{a},\quad 
\overline{C}^{a}\nabla _{\mu }A_{\nu }^{a},\quad \left( \nabla _{\mu }%
\overline{C}^{a}\right) A_{\nu }^{a},\quad \overline{C}^{a}A_{\mu }^{a}, 
\nonumber \\
&&f^{abc}\overline{C}^{a}A_{\mu }^{b}A_{\nu }^{c},\quad f^{abc}\overline{C}%
^{a}\overline{C}^{b}C^{c},\quad K_{\overline{C}}^{a}K_{B}^{a},\quad
f^{abc}K_{B}^{a}A_{\mu }^{b}A_{\nu }^{c},\quad f^{abc}K_{B}^{a}\overline{C}%
^{b}C^{c},  \nonumber \\
&&f^{abc}K_{B}^{a}K_{B}^{b}C^{c},\quad K_{B}^{a}\nabla _{\mu }A_{\nu
}^{a},\quad \left( \nabla _{\mu }K_{B}^{a}\right) A_{\nu }^{a},\quad
K_{B}^{a}A_{\mu }^{a}.  \label{list}
\end{eqnarray}
The counterterms are obtained acting with $\sigma $ on linear combinations
of these objects, but must not contain $B^{a}$, $K_{B}^{a}$ and $K_{%
\overline{C}}^{a}$, because the action (\ref{act}) provides no vertices with 
$B^{a}$, $K_{B}^{a}$ or $K_{\overline{C}}^{a}$ on the external legs. There
are only three linear combinations that have this property, namely 
\begin{equation}
\overline{C}^{a}K_{\overline{C}}^{a}+K_{B}^{a}B^{a}+\lambda \overline{C}%
^{a}B^{a}-\overline{C}^{a}\nabla ^{\mu }A_{\mu }^{a},\qquad \left(
K_{a}^{\mu }+\nabla ^{\mu }\overline{C}^{a}\right) A_{\nu }^{a},\qquad
K_{C}^{a}C^{a}.  \label{three}
\end{equation}
The first of these, however, is trivial, since it is itself $\sigma $-exact.
Precisely, it is equal to $\sigma \left( \overline{C}^{a}K_{B}^{a}\right) $.
In the unperturbed theory, the counterterms are necessarily scalar, so we
can drop the vectors of (\ref{list}) and trace the second of (\ref{three}).
We remain with two scalar combinations. A convenient basis is for example
the one obtained taking the third term of (\ref{three}) and the difference
between the first two, namely 
\begin{equation}
I_{1}(\Phi ,K)\equiv -K_{a}^{\mu }A_{\mu }^{a}+\overline{C}^{a}K_{\overline{C%
}}^{a}+K_{B}^{a}B^{a}+\lambda \overline{C}^{a}B^{a}-\nabla ^{\mu }\left( 
\overline{C}^{a}A_{\mu }^{a}\right) ,\qquad I_{2}(\Phi ,K)\equiv
K_{C}^{a}C^{a}.  \label{abasis}
\end{equation}
In this framework the functional $\mathcal{R}_{n}$ of (\ref{counter}) has
the form 
\[
\mathcal{R}_{n}(\Phi ,K)=\int \mathrm{d}^{n}x\sqrt{-g}\left[ -\delta
_{A}^{(n)}I_{1}+\delta _{C}^{(n)}I_{2}\right] ,
\]
where $\delta _{A}^{(n)}$ and $\delta _{C}^{(n)}$ are divergent constants.
The total derivative contained in the first term of (\ref{abasis}) can be
dropped.

The basis (\ref{abasis}) is such that the $\sigma $-exact counterterms are
reabsorbed by a renormalization $\lambda ^{\prime }=\lambda Z_{n\hspace{%
0.01in}\lambda }$ of the gauge-fixing parameter $\lambda $ and a canonical
transformation 
\[
\Phi ^{i\ \prime }=Z_{n\hspace{0.01in}i}^{1/2}\Phi ^{i},\qquad K_{i}^{(n)\
\prime }=Z_{n\hspace{0.01in}i}^{-1/2}K_{i}, 
\]
generated by 
\[
\mathcal{F}_{n}(\Phi ,K^{\prime })=\int \mathrm{d}^{n}x\sqrt{-g}\sum_{i}Z_{n%
\hspace{0.01in}i}^{1/2}\Phi ^{i}K_{i}^{\prime }=I(\Phi ,K^{\prime })-%
\mathcal{R}_{n}(\Phi ,K^{\prime })+\text{higher orders}. 
\]
where 
\begin{equation}
Z_{n\hspace{0.01in}\overline{C}}=Z_{n\hspace{0.01in}B}=Z_{n\hspace{0.01in}%
A}^{-1}=Z_{n\hspace{0.01in}\lambda }^{-1},\qquad Z_{n\hspace{0.01in}%
K_{i}}=Z_{n\hspace{0.01in}i}^{-1},\qquad Z_{n\hspace{0.01in}%
A}^{1/2}=1-\delta _{A}^{(n)},\qquad Z_{n\hspace{0.01in}C}^{1/2}=1-\delta
_{C}^{(n)}.  \label{warda}
\end{equation}
The Ward identities (\ref{warda}) that relate the wave-function
renormalization constants easily extend to the complete renormalized theory: 
\[
Z_{\overline{C}}=Z_{B}=Z_{A}^{-1}=Z_{\lambda }^{-1},\qquad
Z_{K_{i}}=Z_{i}^{-1},\qquad Z_{i}=\prod_{i=1}^{\infty }Z_{n\hspace{0.01in}%
i}. 
\]

Another parametrization, which is more convenient in the perturbed model,
amounts to use the basis 
\begin{equation}
J_{1}(\Phi ,K)\equiv \left( K_{a}^{\mu }+\nabla ^{\mu }\overline{C}%
^{a}\right) A_{\mu }^{a},\qquad J_{2}(\Phi ,K)\equiv K_{C}^{a}C^{a},
\label{abasis2}
\end{equation}
instead of (\ref{abasis}). Then 
\[
\mathcal{F}_{n}(\Phi ,K^{\prime })=\int \sqrt{-g}\left[ Z_{n\hspace{0.01in}%
A}^{1/2}K_{a}^{\mu \ \prime }A_{\mu }^{a}+\overline{C}^{a}K_{\overline{C}%
}^{a\ \prime }+K_{B}^{a\ \prime }B^{a}+Z_{n\hspace{0.01in}C}^{1/2}K_{C}^{a\
\prime }C^{a}+\left( Z_{n\hspace{0.01in}A}^{1/2}-1\right) \left( \nabla
^{\mu }\overline{C}^{a}\right) A_{\mu }^{a}\right] 
\]
and again $Z_{n\hspace{0.01in}A}^{1/2}=1-\delta _{A}^{(n)}$, $Z_{n\hspace{%
0.01in}C}^{1/2}=1-\delta _{C}^{(n)}$. Here $\overline{C}^{a}$, $B^{a}$, $%
K_{B}^{a}$ and $\lambda $ are non-renormalized and the unique non-trivial
redefinitions are 
\begin{eqnarray}
A_{\mu }^{a} &\rightarrow &Z_{n\hspace{0.01in}A}^{1/2}A_{\mu }^{a},\qquad
K_{a}^{\mu }\rightarrow Z_{n\hspace{0.01in}A}^{-1/2}K_{a}^{\mu }+\nabla
^{\mu }\overline{C}^{a}(Z_{n\hspace{0.01in}A}^{-1/2}-1),\qquad
\label{refeda} \\
K_{\overline{C}}^{a} &\rightarrow &K_{\overline{C}}^{a}-\nabla ^{\mu }A_{\mu
}^{a}+\nabla ^{\mu }(A_{\mu }^{a}Z_{n\hspace{0.01in}A}^{1/2}),\qquad
C^{a}\rightarrow Z_{n\hspace{0.01in}C}^{1/2}C^{a},\qquad
K_{C}^{a}\rightarrow Z_{n\hspace{0.01in}C}^{-1/2}K_{C}^{a},  \nonumber
\end{eqnarray}
besides the renormalization of the gauge coupling.

\bigskip

Now define a map $\Sigma _{\mathcal{L}}$, $\mathcal{L}=(L,L_{A},L_{C})$ made
of a redefinition 
\begin{equation}
\alpha \rightarrow \alpha (fR)=\alpha L(fR)  \label{gauge}
\end{equation}
of the gauge coupling, plus the addition of unspecified pure gravitational
terms (described below), plus the canonical transformation generated by 
\[
\mathcal{F}(\Phi ,K^{\prime })=\int \mathrm{d}^{n}x\sqrt{-g}\left[
L_{A}K_{a}^{\mu \ \prime }A_{\mu }^{a}+\overline{C}^{a}K_{\overline{C}}^{a\
\prime }+K_{B}^{a\ \prime }B^{a}+L_{C}K_{C}^{a\ \prime }C^{a}+\left(
L_{A}-1\right) \left( \nabla ^{\mu }\overline{C}^{a}\right) A_{\mu
}^{a}\right] , 
\]
where the functions $\mathcal{L}$ depend $fR$. The unique non-trivial
redefinitions of fields and BRST\ sources are 
\begin{eqnarray}
A_{\mu }^{a} &\rightarrow &L_{A}A_{\mu }^{a},\qquad K_{a}^{\mu }\rightarrow
L_{A}^{-1}K_{a}^{\mu }+\left( \nabla ^{\mu }\overline{C}^{a}\right)
(L_{A}^{-1}-1),\qquad  \label{refeda2} \\
K_{\overline{C}}^{a} &\rightarrow &K_{\overline{C}}^{a}-\nabla ^{\mu }A_{\mu
}^{a}+\nabla ^{\mu }(A_{\mu }^{a}L_{A}),\qquad C^{a}\rightarrow
L_{C}C^{a},\qquad K_{C}^{a}\rightarrow L_{C}^{-1}K_{C}^{a},  \nonumber
\end{eqnarray}

The maps $\Sigma _{\mathcal{L}}$ form a group: 
\[
\Sigma _{\mathcal{L}}\Sigma _{\mathcal{L}^{\prime }}=\Sigma _{\mathcal{LL}%
^{\prime }},\qquad \mathcal{LL}^{\prime }=(LL^{\prime },L_{A}L_{A}^{\prime
},L_{C}L_{C}^{\prime }). 
\]
Moreover, the basis (\ref{abasis2}) is invariant: 
\[
\Sigma _{\mathcal{L}}J_{1}=J_{1},\qquad \Sigma _{\mathcal{L}}J_{2}=J_{2}. 
\]

We assert that the perturbed theory obtained applying the map $\Sigma _{%
\mathcal{L}}$ to the unperturbed one is renormalizable and that the
subtraction of divergences is again a map $\Sigma _{\mathcal{L}}$, namely a
renormalization of the gauge coupling of the form (\ref{gauge}), plus a
canonical transformation of the form (\ref{refeda2}), plus a suitable
renormalization of the pure gravitational terms.

\bigskip

The action of the perturbed theory is 
\begin{eqnarray}
\mathcal{S}_{\text{EYM-HD}}^{(f)} &=&\Sigma _{\mathcal{L}}\mathcal{S}_{\text{%
EYM-HD}}=\int \mathrm{d}^{n}x\sqrt{-g}\left\{ \frac{1}{2\kappa ^{2}}R-\frac{1%
}{4\alpha L}F_{\mu \nu }^{a}(L_{A}A)F^{a\hspace{0.01in}\mu \nu }(L_{A}A)-%
\frac{\lambda }{2}B^{a}B^{a}+B^{a}\nabla ^{\mu }A_{\mu }^{a}\right. 
\nonumber \\
&&\!\!\!\!\!\!\!\!\!\!\!\!\!\!\!\!\!\!\!\!\!\!\!\!\!\!\!\left.
+L_{A}^{-1}\left( K_{a}^{\mu }+\nabla ^{\mu }\overline{C}^{a}\right) \left[
\partial _{\mu }(L_{C}C^{a})+f^{abc}L_{A}L_{C}A_{\mu }^{b}C^{c}\right] +%
\frac{L_{C}}{2}f^{abc}K_{C}^{a}C^{b}C^{c}-K_{\overline{C}}^{a}B^{a}\right\}
+\Delta S_{g},  \label{th}
\end{eqnarray}
where $\Delta _{g}S$ denote the pure gravitational terms, so far
unspecified. The Batalin-Vilkovisky analysis has to be applied with the
perturbed $\sigma $-operator $\sigma _{f}$, defined by 
\[
\sigma _{f}X\equiv \left( \mathcal{S}_{\text{EYM-HD}}^{(f)},X\right) . 
\]
It is easy to verify that $\sigma _{f}$ is nilpotent, namely 
\[
\left( \mathcal{S}_{\text{EYM-HD}}^{(f)},\mathcal{S}_{\text{EYM-HD}%
}^{(f)}\right) =0. 
\]
Calling $\Gamma _{\text{EYM-HD}}^{(f)}$ the generating functional of
one-particle irreducible diagrams, we have also 
\begin{equation}
(\Gamma _{\text{EYM-HD}}^{(f)},\Gamma _{\text{EYM-HD}}^{(f)})=0.
\label{milpo2}
\end{equation}
Proceeding inductively, call $\Gamma _{\text{EYM-HD}}^{\ (f,n)}$ the
perturbed generating functional renormalized up to the $n$-th loop included.
Renormalization theory and (\ref{milpo2}) ensure that the ($n+1$)-loop
divergences $\Gamma _{\text{div}}^{\ (f,n+1)}$ are local and $\sigma _{f}$%
-closed: $\sigma _{f}\Gamma _{\text{div}}^{\ (f,n+1)}=0$. The solution is 
\begin{equation}
\Gamma _{\text{div}}^{\ (f,n+1)}=\mathcal{G}_{n\hspace{0.01in}f}+\sigma _{f}%
\mathcal{R}_{n\hspace{0.01in}f}.  \label{sola}
\end{equation}

Let us analyze the vertices of the theory, that can be read from (\ref{th}).
We can write 
\[
\mathcal{S}_{\text{EYM-HD}}^{(f)}=\mathcal{S}_{\text{EYM-HD}}+\int \mathrm{d}%
^{n}x\sqrt{-g}\left\{ \overline{K}_{1}\mathrm{O}_{1}+\overline{K}_{2\hspace{%
0.01in}\mu }\mathrm{O}_{2}^{\mu }+\overline{K}_{3\hspace{0.01in}\mu \nu }%
\mathrm{O}_{3}^{\mu \nu }\right\} +\Delta S_{g}, 
\]
where $\mathrm{O}_{1}$, $\mathrm{O}_{2}^{\mu }$ and $\mathrm{O}_{3}^{\mu \nu
}$ are $f$-independent operators, with dimensionalities 4, 3 and 2,
respectively, constructed with the fields, the BRST\ sources and their
derivatives, while the gravitational sources read 
\begin{equation}
\overline{K}_{1}=P_{1}(fR),\qquad \overline{K}_{2\hspace{0.01in}\mu
}=fP_{2}(fR)\nabla _{\mu }R,\qquad \overline{K}_{3\hspace{0.01in}\mu \nu
}=f^{2}P_{3}(fR)\nabla _{\mu }R\nabla _{\nu }R.  \label{list2}
\end{equation}
the $P_{i}$'s being functions of $fR$. Every $f$-dependence is contained in
the $\overline{K}$'s.

The counterterms (\ref{sola}) are local, covariant, have dimensionality four
and are constructed with the $\overline{K}$'s, the matter fields $\Phi ^{i}$%
, the BRST\ sources $K_{i}$, the curvature tensors and their covariant
derivatives. The $\overline{K}$'s have non-negative dimensionalities in
units of mass. There is one ingredient ($P_{1}$) of dimensionality zero,
which is why renormalization turns on arbitrary functions of $fR$.

Let us study the $\sigma _{f}$-cohomology. We postpone the discussion about
the pure gravitational terms, which are trivially $\sigma _{f}$-closed, and
first concentrate on the terms that depend on $\Phi ^{i}$ and $K_{i}$. The $%
\sigma _{f}$-closed terms of type $\mathcal{G}_{n\hspace{0.01in}f}$ can
contain $F^{2}$, $T_{\mu \nu }$ and $\Upsilon _{\mu \nu \rho \sigma }$, with 
$A_{\mu }^{a}$ replaced by $L_{A}A_{\mu }^{a}$. However, power counting
excludes both $T_{\mu \nu }$ and $\Upsilon _{\mu \nu \rho \sigma }$, since
they have dimensionality four and the only dimensionless $\overline{K}$ is
scalar. Therefore, only $F^{2}$ remains.

The functional $\mathcal{R}_{n\hspace{0.01in}f}$ appearing in the exact $%
\sigma _{f}$-terms of (\ref{sola}) is again a linear combination of (\ref
{list}), with coefficients constructed with the sources $\overline{K}$'s,
the curvature tensors and their covariant derivatives, such that $\sigma _{f}%
\mathcal{R}_{n\hspace{0.01in}f}$ does not contain $B^{a}$, $K_{B}^{a}$ and $%
K_{\overline{C}}^{a}$. There are no $\sigma _{f}$-exact terms with
dimensionality two or less, so $\overline{K}_{3\hspace{0.01in}\mu \nu }$ can
be dropped. We can drop also $\overline{K}_{2\hspace{0.01in}\mu }$ together
with the terms $\overline{C}^{a}A_{\mu }^{a}$ and $K_{B}^{a}A_{\mu }^{a}$ of
(\ref{list}), because the counterterms constructed with these objects are
easily converted, by means of a partial integration, into products of a
scalar function times a combination of the other terms (\ref{list}).

The $\sigma $- and $\sigma _{f}$- cohomologies are in one-to-one
correspondence. To relate them, start from 
\begin{equation}
\sigma X=(\mathcal{S}_{\text{EYM-HD}},X)\equiv Y  \label{sx}
\end{equation}
and perform the canonical transformation (\ref{refeda2}). Using the
invariance of the antiparenthesis and denoting the transformed functionals
with a tilde, we obtain 
\begin{equation}
(\widetilde{\mathcal{S}}_{\text{EYM-HD}},\widetilde{X})=\widetilde{Y}.
\label{sx2}
\end{equation}
The transformed action $\widetilde{\mathcal{S}}_{\text{EYM-HD}}$ differs
from $\mathcal{S}_{\text{EYM-HD}}^{(f)}$ because of the coupling
redefinition: 
\[
\widetilde{\mathcal{S}}_{\text{EYM-HD}}=\mathcal{S}_{\text{EYM-HD}%
}^{(f)}+\int \mathrm{d}^{n}x\sqrt{-g}\frac{1-L}{4\alpha L}F_{\mu \nu
}^{a}(L_{A}A)F^{a\hspace{0.01in}\mu \nu }(L_{A}A)\equiv \mathcal{S}_{\text{%
EYM-HD}}^{(f)}+\widetilde{\Delta }_{L}, 
\]
therefore (\ref{sx2}) can be written as 
\[
\sigma _{f}\widetilde{X}=\widetilde{Y}-(\widetilde{\Delta }_{L},\widetilde{X}%
)=\widetilde{Y}-\widetilde{(\Delta _{L},X)}. 
\]

Assume that $X$ is a linear combination of the terms (\ref{list}). Since $%
\Delta _{L}$ depends only on $A_{\mu }^{a}$, only the term proportional to $%
K_{a}^{\mu }A_{\nu }^{a}$ in $X$ contributes to $(\Delta _{L},X)$. So, $%
(\Delta _{L},X)$ does not contain $B^{a}$, $K_{B}^{a}$ or $K_{\overline{C}%
}^{a}$. Moreover, the canonical transformation (\ref{refeda2}) is such that
functionals (not) containing $B^{a}$, $K_{B}^{a}$ and $K_{\overline{C}}^{a}$
are mapped into functionals (not) containing $B^{a}$, $K_{B}^{a}$ and $K_{%
\overline{C}}^{a}$. Thus, an $X$ such that $\sigma X$ does (not) contain $%
B^{a}$, $K_{B}^{a}$ and $K_{\overline{C}}^{a}$ is mapped into an $\widetilde{%
X}$ such that $\sigma _{f}\widetilde{X}$ does (not) contain $B^{a}$, $%
K_{B}^{a}$ and $K_{\overline{C}}^{a}$, and viceversa. Having dropped both $%
\overline{K}_{2\hspace{0.01in}\mu }$ and $\overline{K}_{3\hspace{0.01in}\mu
\nu }$, we can focus on scalar functionals $X$, $\widetilde{X}$.

These properties ensure that most general $\widetilde{X}$ can be obtained
applying the canonical transformation (\ref{refeda2}) to the most general $X$%
. Since the latter is a linear combination of $J_{1}$ and $J_{2}$, and $%
\widetilde{J}_{1}=J_{1}$, $\widetilde{J}_{2}=J_{2}$, also the former is a
linear combination of $J_{1}$ and $J_{2}$.

Finally, we have 
\[
\Gamma _{\text{div}}^{(f,n+1)}=U_{n}\ F_{\mu \nu }^{a}(L_{A}A)F^{a\mu \nu
}(L_{A}A)+\sigma _{f}(V_{n}J_{1}+W_{n}J_{2})+\text{ pure gravitational terms,%
} 
\]
where $U_{n}$, $V_{n}$ and $W_{n}$ are functions of $fR$. Now it is
immediate to see that the divergences are inductively subtracted by a map of
the form $\Sigma _{\mathcal{L}}$ with $\mathcal{L}=(1-4\alpha
LU_{n},1-V_{n},1-W_{n})$ and we conclude that the theory (\ref{th}) is
renormalizable.

The pure gravitational counterterms can be constructed with the $\overline{K}
$'s, the curvature tensors and their covariant derivatives. The list of
independent terms is 
\begin{eqnarray}
&&Q_{1}R_{\mu \nu \rho \sigma }R^{\mu \nu \rho \sigma },\qquad Q_{2}R_{\mu
\nu }R^{\mu \nu },\qquad Q_{3}R^{2},\qquad Q_{4}\Box R,\qquad
f^{2}Q_{5}R^{\mu \nu }\nabla _{\mu }R\nabla _{\nu }R,  \label{eight} \\
&&f^{3}Q_{6}\nabla ^{\mu }R\nabla _{\mu }R\Box R,\qquad f^{4}Q_{7}\left(
\nabla ^{\mu }R\nabla _{\mu }R\right) ^{2},\qquad f^{2}Q_{8}\left( \Box
R\right) ^{2},  \nonumber
\end{eqnarray}
where $Q_{i}$, $i=1,\ldots 8$ are functions of $fR$. Thus $\Delta _{g}S$ is
linear combination of such terms. We see that there is only one vertex, $%
RR_{\mu \nu \rho \sigma }R^{\mu \nu \rho \sigma }$, that is not squarely
proportional to the Ricci tensor.

\end{document}